\documentclass[
  reprint,
  superscriptaddress,
  amsmath,amssymb,
  aps,
  prb,
  nofootinbib,
  floatfix,
  longbibliography
]{revtex4-2}

\usepackage[T1]{fontenc}
\usepackage[utf8]{inputenc}
\usepackage{lmodern}
\usepackage{textcase}

\usepackage{amsmath,amssymb,bm,mathtools}
\usepackage{physics}
\usepackage{siunitx}
\usepackage{float} 

\usepackage{graphicx}
\usepackage{xcolor}
\usepackage{dcolumn}
\usepackage{booktabs}

\usepackage[colorlinks=true,allcolors=blue]{hyperref}

\newcommand{\TN}{T_{\mathrm{N}}}

\makeatletter
\def\frontmatter@preabstractspace{12pt}
\renewcommand{\fnum@figure}{\textbf{Fig.~\thefigure}}
\renewcommand{\fnum@table}{\textbf{Table~\thetable}}
\makeatother

\begin{document}

\title{Raman signatures of a non-reciprocal magnetic phase transition in Ca$_2$RuO$_4$}

\author{Giacomo Jarc}
\thanks{These authors contributed equally to this work.}
\affiliation{Department of Physics, University of Erlangen-Nürnberg, Erlangen, Germany}

\author{Giovanni Tartaglia}
\thanks{These authors contributed equally to this work.}
\affiliation{Department of Physics, University of Trieste, Trieste, Italy}

\author{Francesco Gabriele}
\affiliation{CNR-SPIN, University of Salerno, Fisciano, Salerno, Italy}

\author{Filomena Forte}
\affiliation{CNR-SPIN, University of Salerno, Fisciano, Salerno, Italy}

\author{Anita Guarino}
\affiliation{CNR-SPIN, University of Salerno, Fisciano, Salerno, Italy}

\author{Angela Montanaro}
\affiliation{Department of Physics, University of Erlangen-Nürnberg, Erlangen, Germany}

\author{Enrico Maria Rigoni}
\affiliation{Department of Physics, University of Erlangen-Nürnberg, Erlangen, Germany}

\author{Nitesh Khatiwada}
\affiliation{Department of Physics, University of Erlangen-Nürnberg, Erlangen, Germany}

\author{Costanza Lincetto}
\affiliation{Department of Physics, University of Erlangen-Nürnberg, Erlangen, Germany}

\author{Gabriele Bartolini}
\affiliation{Department of Physics, University of Erlangen-Nürnberg, Erlangen, Germany}

\author{Antonio Mastropasqua}
\affiliation{Department of Physics, University of Erlangen-Nürnberg, Erlangen, Germany}

\author{Shahla Yasmin Mathengattil}
\affiliation{Department of Physics, University of Trieste, Trieste, Italy}
\affiliation{Elettra Sincrotrone Trieste, Trieste, Italy}

\author{Marco Malvestuto}
\affiliation{Elettra Sincrotrone Trieste, Trieste, Italy}
\affiliation{CNR-Istituto Officina dei Materiali (IOM), Trieste, Italy}

\author{Muhammad Waqee Ur Rehman}
\affiliation{CNR-SPIN, University of Salerno, Fisciano, Salerno, Italy}

\author{Rosalba Fittipaldi}
\affiliation{CNR-SPIN, University of Salerno, Fisciano, Salerno, Italy}

\author{Joachim Deisenhofer}
\affiliation{Experimental Physics V, Center for Electronic
Correlations and Magnetism, Institute for Physics, University of Augsburg, D-86159 Augsburg, Germany}

\author{Alexander A. Tsirlin}
\affiliation{Felix Bloch Institute for Solid-State Physics, Leipzig University, 04103 Leipzig, Germany}

\author{Antonio Vecchione}
\affiliation{CNR-SPIN, University of Salerno, Fisciano, Salerno, Italy}

\author{Mario Cuoco}
\affiliation{CNR-SPIN, University of Salerno, Fisciano, Salerno, Italy}

\author{Daniele Fausti}
\thanks{Corresponding author: \href{mailto:daniele.fausti@fau.de}{daniele.fausti@fau.de}}
\affiliation{Department of Physics, University of Erlangen-Nürnberg, Erlangen, Germany}

\begin{abstract}
The magnetic behavior of Ca$_2$RuO$_4$ represents a unique puzzle due to the interplay of strong electronic correlations, magneto-elastic interactions, and large spin-orbit coupling. At low temperatures, an anomalous Mott insulating state emerges, characterized by a complex antiferromagnetic order with a collective amplitude excitation of the magnetic moment which has been discussed in analogy with the Higgs mode. We report here evidence of a magnetic first-order phase transition driven by an out-of-plane magnetic field along the crystallographic $c$-axis of Ca$_2$RuO$_4$. Raman measurements in magnetic field reveal the emergence of mode of magnetic origin and a concomitant modification of the coupling between a phonon and the Higgs amplitude mode. Both the Raman features are characterized by a non-reciprocal hysteresis in magnetic field. Surprisingly, the observed phase transition does not affect the Raman scattering from the in-plane magnons, indicating that the dipolar antiferromagnetic order is preserved. This is consistent with the onset of a new quadrupolar field-controlled state, whose fluctuations structure can trigger an hybridization between the lattice and the magnetic Higgs mode resulting in the observed Raman features.
\end{abstract}

\maketitle

\section{Introduction}

\begin{figure*}[t]
\centering
\includegraphics[width=0.76\textwidth]{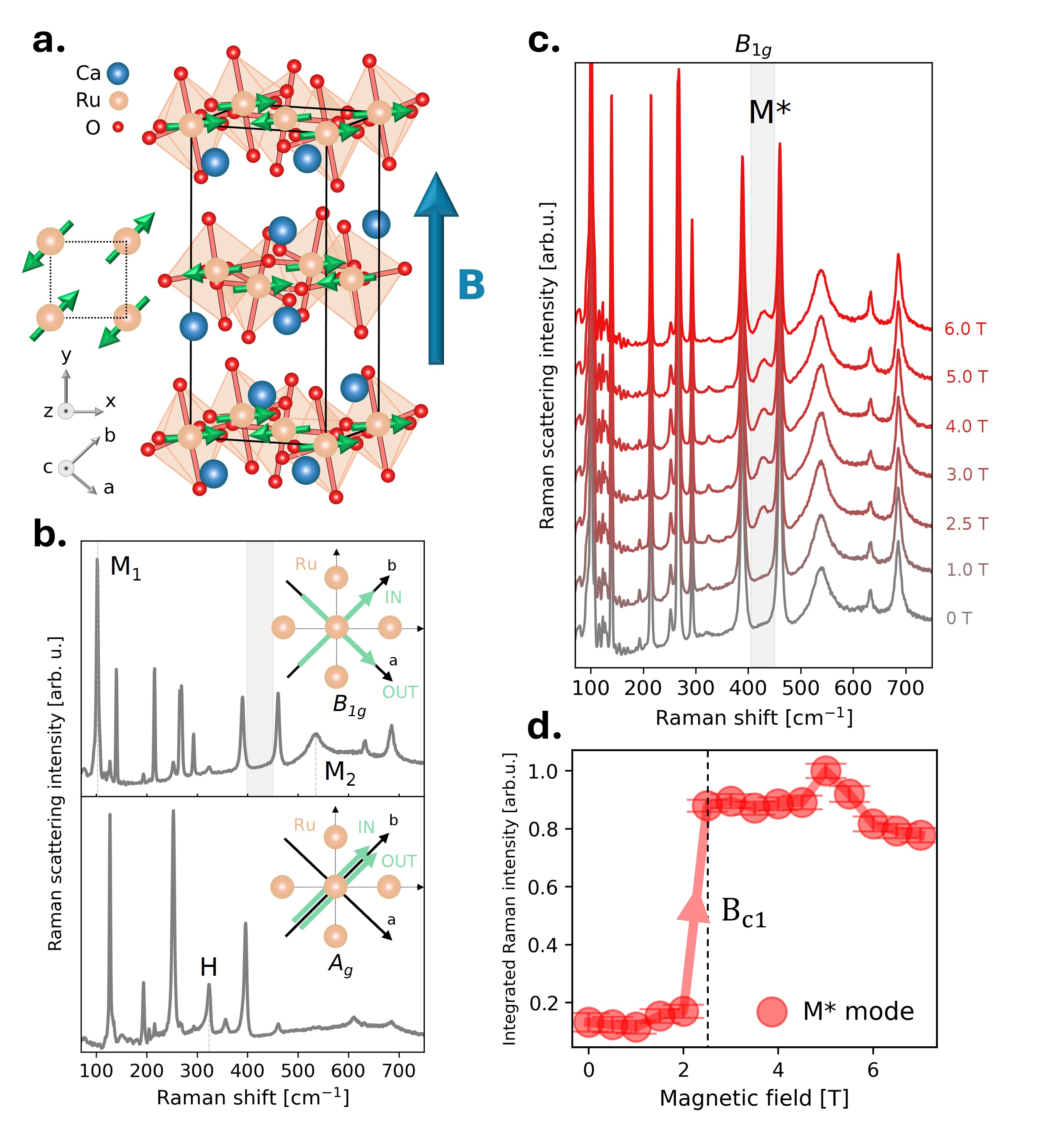}
\caption{
\textbf{Emergence of a Raman mode in Ca$_2$RuO$_4$ in $c$-axis magnetic field.}
\textbf{(a)} Crystal structure and schematic spin configuration of Ca$_2$RuO$_4$. A magnetic field is applied across the antiferromagnetic (AF) planes, along the $c$-axis ($B \parallel c$). \textbf{(b)} Raman scattering from the AF phase of Ca$_2$RuO$_4$ at $1.6$ K in $B_{1g}$ and $A_g$ symmetries. The $M_1$ and $M_2$ peaks denote the magnon modes of the in-plane AF order, while the $H$ peak in $A_g$ symmetry denotes the strongly asymmetric phonon coupled to the Higgs mode. Insets show the scattering geometries of the $B_{1g}$ and $A_g$ symmetries, illustrating the polarization configurations relative to the $Pbca$ crystal axes. \textbf{(c)} Magnetic-field-dependent Raman spectra in $B_{1g}$ symmetry for fields from $0$ to $6$ T, showing the emergence of the $M^*$ mode. Spectra are offset vertically for clarity. \textbf{(d)} Integrated Raman scattering of the $M^*$ mode (432 cm$^{-1}$ central frequency, 20 cm$^{-1}$ window) as a function of the magnetic field, revealing a critical field $B_{c1} \approx 2.5$ T. The integral is calculated as the sum of $B_{1g}$ and $A_g$ channels for each magnetic field to avoid Faraday rotation artifacts. Integrals of $M^*$ are normalized to the maximum.}
\label{fig1}
\end{figure*}

Ca$_2$RuO$_4$ is a $4d^4$ Mott insulator in which lattice, charge, orbital, and spin degrees of freedom are strongly intertwined, and give rise to a peculiar electronic and magnetic behavior ~\cite{nakatsuji1997, gretarsson2019, jung2003, cuono2025,Gauquelin2023,Curcio2023}. At $T_{\mathrm{MI}} \sim 360$ K~\cite{alexander1999, friedt2001} Ca$_2$RuO$_4$ undergoes a metal insulator transition which is associated to structural distortions of the RuO$_6$ octahedra. The rotation, tilt, and flattening of the octahedra by cooling results in a narrowing of the Ru$^{4+}$ $t_{2g}^4$ bands which in turns lead to an insulating behavior. Upon further cooling, an antiferromagnetic (AF) Mott phase emerges below the N\'eel temperature ($T_{\mathrm{N}} \sim 110$ K)~\cite{braden1998, gorelov2010}. In this phase, the Ru$^{4+}$ magnetic moments are predominantly aligned along the crystallographic $b$-axis~\cite{mizokawa2001, carlo2012}, with a small canting toward the $c$-axis~\cite{porter2018}. 

The energy proximity between the $J = 0$ and $J = 1$ spin-orbit states of Ru$^{4+}$ $t_{2g}^4$ manifold, gives to the antiferromagnetic order in Ca$_2$RuO$_4$ the peculiar character excitonic-magnetisim highly discussed in recent literature ~\cite{khaliullin2013, akbari2014, jain2017, fatuzzo2015, das2018, vonArx2025, feldmaier2020,
Bhartiya2025}. Furthermore, the $S=1$ local Hilbert space of Ru$^{4+}$ ions in Ca$_2$RuO$_4$ naturally supports magnetic degrees of freedom beyond the conventional dipolar (rank-1) moment, including higher-rank multipoles such as quadrupolar (rank-2) order. In essence, multipolar degrees of freedom allow the system to order in the shape of its spin fluctuations rather than their average direction, so that distinct ordered states with fundamentally different fluctuation structures can coexist with the same (vanishing) expectation values of ⟨S⟩. Such multipolar states have been investigated both in Ca$_2$RuO$_4$ and in other correlated materials \cite{Santini2009,Pourovskii2021}, underscoring their broader relevance in strongly spin-orbit-coupled systems~\cite{nag2023, ning2023, laeuchli2006, lu2017, kim2024,Mazzola2024,Mazzola2026}.

In this work, we investigate the response of $\text{Ca}_2\text{Ru}\text{O}_4$ to an external magnetic field applied along the crystallographic $c$-axis ($B \parallel c$). While in-plane magnetic fields are known to drive a metamagnetic spin-flop transition~\cite{cao1997}, the effect of an out-of-plane field on the magnetic ground state remains unexplored. Rather than relying on static perturbations such as substitutional doping, which alter the intralayer magnetic exchange~\cite{porter2022,pincini2019,Brzezicki2023}, we employ a $c$-axis magnetic field as a tunable parameter to control the magnetic ground state of $\text{Ca}_2\text{Ru}\text{O}_4$. Along the $c$-axis, the magnetic field directly couples with the intrinsic out-of-plane spin canting and the staggered magnetization of the Ru$^{4+}$ ions. Given the $S=1$ nature of the AF phase in Ca$_2$RuO$_4$, the modification of its out-of-plane magnetism can activate higher-order degrees of freedom of the local magnetic moments, triggering the emergence of a quadrupolar phase.

To investigate the impact of a $c$-axis magnetic field on the antiferromagnetic ground state of Ca$_2$RuO$_4$, we exploit polarization-resolved Raman scattering. Raman spectroscopy is known to be sensitive to collective excitations in ruthenates~\cite{rho2003, rho2005, Wulferding2026} and importantly provides access to magnetic excitations in the form of single- and multi-magnon modes, whose energies reflect the long-range antiferromagnetic order of the Ru$^{4+}$ spins ~\cite{kunkemoller2015}. Additionally, by simultaneously probing structural phonons, Raman scattering offers a direct view of the magneto-elastic coupling in the AF ground state~\cite{rho2005, lee2019}. Exploiting this sensitivity of Raman scattering to collective structural and magnetic excitations, we report evidence of a first-order magnetic phase transition induced by a $c$-axis magnetic field in Ca$_2$RuO$_4$. Unlike the metamagnetic transition driven by an in-plane magnetic field~\cite{cao1997}, which alters the in-plane magnetic order and rearranges its collective magnon excitations, the transition observed for $B \parallel c$ leaves the primary in-plane antiferromagnetic order preserved. This is demonstrated by the fact that the magnon scattering processes remain unaltered across the magnetic transition. The Raman response of Ca$_2$RuO$_4$ in a $c$-axis magnetic field reveals instead a selective reconstruction of the AF ground state, characterized by the emergence of a new magnetic mode and the modification of the Fano anomaly of a phonon coupled to the magnetic Higgs excitation.

To account for these observations, we develop a phenomenological Ginzburg–Landau free-energy description in which a quadrupolar moment serves as the primary order parameter. Within this framework, the field-induced onset of quadrupolar order modifies the magneto-elastic coupling and promotes a Higgs–phonon hybridization, consistent with both the emergence of the new magnetic excitation and the evolution of the phonon's Fano anomaly. A key prediction of the model is the occurrence of a non-reciprocal hysteresis associated with the quadrupolar order parameter across the first-order transition.

Experimentally, both the emergent magnetic mode and the correlated Fano anomaly exhibit such non-reciprocal hysteresis as a function of the applied out-of-plane magnetic field. The observation of this anomalous hysteretic behavior, together with the preservation of the primary dipolar antiferromagnetic order, indicates the activation of a field-induced quadrupolar sector of the magnetic moment distribution. The correspondence between the measured Raman hysteresis and the prediction of the Ginzburg–Landau description supports the identification of quadrupolar order as the driving degree of freedom of the magnetic transition.

\section{Experimental Methods}

Single crystals of Ca$_2$RuO$_4$ are cleaved such that the incident light direction is perpendicular to the $ab$ plane. The magnetic field is applied along the $c$-axis ($B \parallel c$). Raman scattering is measured down to 1.6 K in a backscattering geometry using a 532 nm continuous-wave (CW) laser, with an incident power of 3~mW focused onto a $\sim$10~$\mu$m spot size, through a long-working-distance microscope objective (50$\times$ magnification, numerical aperture $\text{NA} = 0.7$, working distance $7$ mm). The scattered light is dispersed by a single-grating spectrometer (1200 grooves/mm), providing a spectral resolution of $\sim 0.3$ cm$^{-1}$.

Raman spectra are acquired using linearly polarized light, which is controlled in both input and output to isolate the $B_{1g}$ and $A_g$ symmetries.

The magnetic field protocol consists of a full magnetic cycle: an increasing sweep (0 T to +7 T), a decreasing sweep (+6.5 T to -6.5 T), and a closing sweep (-6.5 T to 0 T).
To avoid systematic artifacts from Faraday rotation in the microscope objective under magnetic field, we perform an independent calibration of the incoming and outgoing polarizations on a non-magnetic Sr$_2$RuO$_4$ reference sample (see SM, Figure \ref{fig:S7} for further details).

\begin{figure*}[t]
\centering
\includegraphics[width=\textwidth]{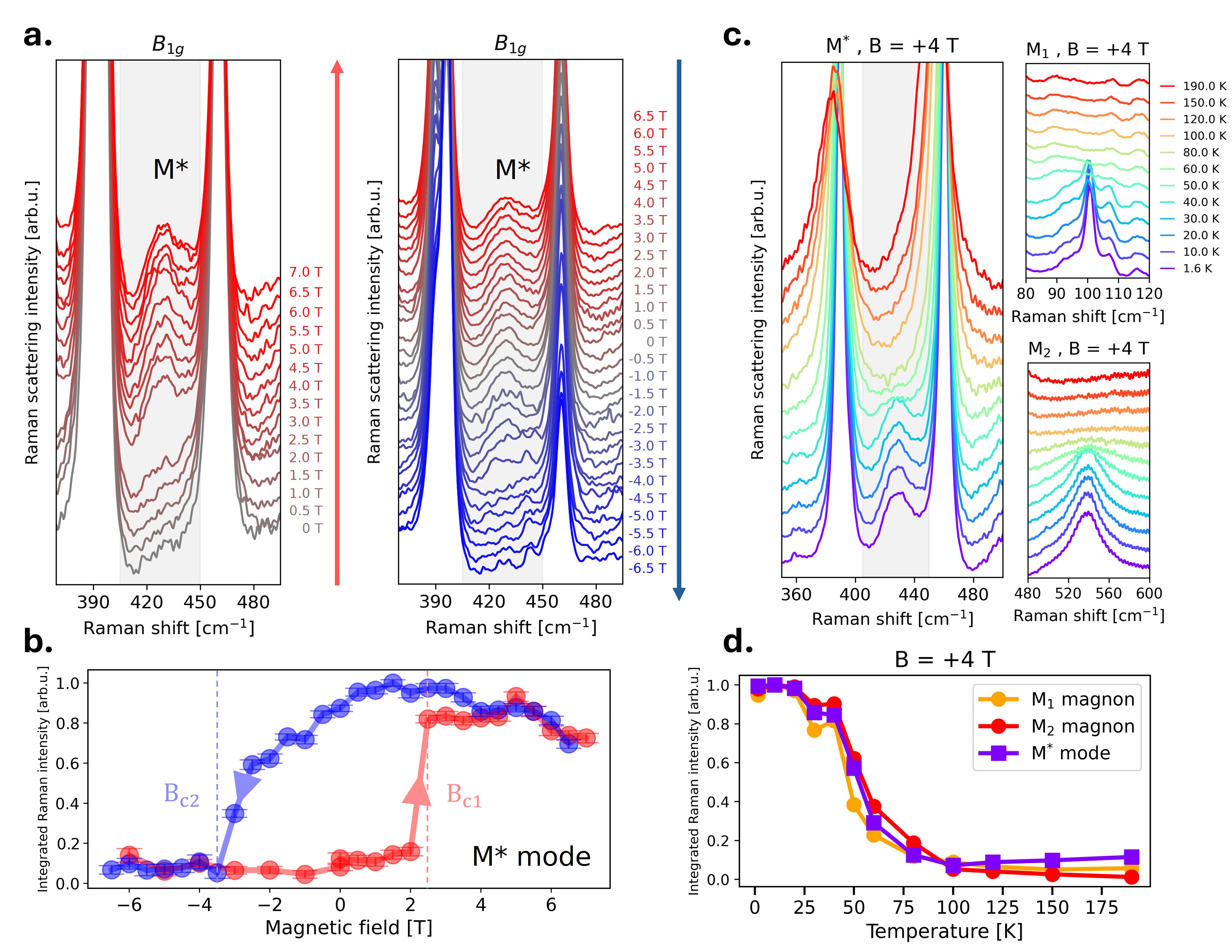}
\caption{ \textbf{Magnetic-field hysteresis and temperature dependence of the $M^*$ Raman mode.} \textbf{(a)} Raman spectra in $B_{1g}$ symmetry at T$=1.6$ K for increasing magnetic field from $0$ T to $+7$ T and decreasing field from $+6.5$ T to $-6.5$ T. The mode $M^*$ emerging in magnetic field is highlighted in the shaded region. \textbf{(b)} Integrated intensity of $M^*$ (evaluated within a 20 cm$^{-1}$ window around 432 cm$^{-1}$) versus field (calculated as the sum of $B_{1g}$ and $A_g$ channels to avoid Faraday rotation artifacts), revealing a non-reciprocal hysteresis with switching fields $B_{c1} \approx +2.5$ T and $B_{c2} \approx -3.5$ T. $M^*$ is the sole mode exhibiting this hysteresis. The integrated $M^*$ intensities are normalized to their maximum value across the entire magnetic-field range. \textbf{(c)} Temperature-dependent Raman spectra at fixed field $B=+4$ T; the left side highlights the evolution of $M^*$ mode, while the right side shows the magnon modes $M_1$ and $M_2$. \textbf{(d)} Normalized integrated Raman scattering of $M^*$, together with the one of $M_1$ and $M_2$, at $B=+4$ T. The temperature evolution of $M^*$ follows the behavior of the $M_1$ and $M_2$ magnons, indicating that $M^*$ is linked to the antiferromagnetic phase.}
\label{fig2}
\end{figure*}

\section{Emergence of a hysteretic magnetic Raman mode in magnetic field}

Figure \ref{fig1}(a) illustrates the orthorhombic crystal structure of Ca$_2$RuO$_4$, highlighting the RuO$_6$ octahedra and the spin configuration of the antiferromagnetic ground state. In the experiment, the external magnetic field is applied along the crystallographic $c$-axis ($B \parallel c$), perpendicular to the primary antiferromagnetic $ab$ planes. This configuration enables direct coupling of the magnetic field with the out-of-plane spin canting.

In the insulating phase, Ca$_2$RuO$_4$ crystallizes in the orthorhombic $Pbca$ space group (point group $D_{2h}$, $G=mmm$) with a $k=0$ antiferromagnetic order~\cite{porter2018}, where the magnetic unit cell matches the paramagnetic one. The $Pbca$ lattice symmetry defines the Raman selection rules, enabling the isolation of distinct magnetic and structural excitations within the $A_g$ and $B_{1g}$ channels in in-plane backscattering geometry~\cite{rho2003, rho2005}. 

Figure \ref{fig1}(b) presents the static Raman spectra measured at 1.6 K and zero magnetic field. As illustrated in the insets of Figure \ref{fig1}(b), the sample is probed at $45^\circ$ with respect to the Ru$^{4+}$ sublattice of the AF planes, with the incident light polarization aligned along the $b$-axis (spin direction). This results in the parallel $c(bb)\bar{c}$ polarization geometry for the $A_g$ channel and the extinction $c(ba)\bar{c}$ for the $B_{1g}$ channel.

Specifically, the $B_{1g}$ channel isolates the magnetic spin-flip excitations, featuring the sharp single-magnon mode $M_1$ at $\sim 100$ cm$^{-1}$ and the broader two-magnon continuum $M_2$ centered at $\sim 538$ cm$^{-1}$~\cite{souliou2017}. The $A_g$ spectrum reveals multiple structural phonons superimposed on a broad continuum background of magnetic origin. DFT calculations of the expected phonon modes at $1.6$~K for both symmetries are provided in Section \ref{section:S2} of the SM.

In this work, we focus on the dynamics of the $H$ phonon mode at $324\,\mathrm{cm}^{-1}$, which corresponds to the internal rotation of the RuO$_6$ octahedra leading to a straightening of their relative orientation. A sketch of the atomic displacement corresponding to the $H$ mode is shown in Figure ~\ref{fig:supp_Hmode} of the SM. This mode exhibits a pronounced Fano asymmetry due to its coupling with the underlying magnetic continuum. As detailed by the temperature-dependent analysis in the SM (Figure \ref{S1}), this magnetic background is composite. It comprises the magnetic Higgs amplitude mode of the spin-orbit condensate, which peaks near the $H$-mode frequency~\cite{souliou2017} and vanishes above $T_{\mathrm{N}}$, superimposed on a broader continuum of short-range antiferromagnetic correlations that is also present in the $B_{1g}$ channel. Consequently, as the system warms up, the magnitude of the Fano asymmetry parameter of the $H$-mode ($q_{\mathrm{F}}$) starts to decrease around $T_{\mathrm{N}}$, reflecting the collapse of the Higgs mode. However, because the short-range magnetic fluctuations survive deep into the paramagnetic phase, they continue to provide an active scattering continuum for the $H$ phonon, sustaining a non-zero $q_{\mathrm{F}}$ also above the transition (see SM, Figure \ref{S1}).

Figure \ref{fig1}(c) displays the magnetic field-dependent Raman spectra in the $B_{1g}$ symmetry for selected field values up to 6 T. Upon increasing the magnetic field, a new Raman mode, $M^*$, emerges. This mode is absent at zero field and is characterized by a central frequency of approximately 432 cm$^{-1}$ and a broad linewidth of $\sim 15$ cm$^{-1}$.

Figure \ref{fig1}(d) tracks the integrated intensity of the $M^*$ mode (evaluated within a 20 cm$^{-1}$ spectral window) as a function of the magnetic field. The activation of the mode occurs above a critical threshold $B_{c1} \approx 2.5$ T. To further ensure the Raman scattering intensity to be independent of Faraday-induced polarization rotation in the microscope objective, the integrals plotted in Figure \ref{fig1}(d) are calculated as the sum of the $B_{1g}$ and $A_g$ components ($I_{Tot} = I_{B_{1g}} + I_{A_g}$).
Crucially, the absence of any mode at $\sim 432$ cm$^{-1}$ in the $A_g$ channel further rules out polarization leakage due to Faraday effect, demonstrating that $M^*$ is a new excitation of $B_{1g}$ symmetry (see SM, Figure \ref{fig:S6}). The field-induced origin of $M^*$ is confirmed by DFT calculations, which show that no phonon mode is present at this frequency in the zero-field AF state for either symmetry (see Section \ref{section:S2} of the SM, Table \ref{tab:eigenfrequencies}).

While $M^*$ grows under the out-of-plane magnetic field, all other $B_{1g}$ excitations observed in zero field, namely the $M_1$ and $M_2$ magnons and the lattice phonons, remain almost invariant in energy and cross section (see Figures \ref{S2}, \ref{S3}, and \ref{fig:S8} in the SM for detailed analysis). Crucially, the invariance of both the single magnon $M_1$ and the two-magnon $M_2$ reveals that the external magnetic field activates a specific excitation without perturbing the in-plane N\'eel vector of the Ru$^{4+}$ spins.

To elucidate the nature of the emergent $M^*$ mode and its coupling to the out-of-plane magnetic field, we perform a magnetic-field loop extending from $B=-6.5$ T to $B=+7$ T at 1.6 K and investigate its temperature dependence across the AF transition.

Figure \ref{fig2}(a) presents the evolution of the $B_{1g}$ Raman spectra acquired during a full magnetic-field sweep, highlighting the frequency region of the $M^*$ Raman mode. We observe a hysteresis in the $M^*$ mode's intensity that is non-reciprocal with respect to the field direction. As shown in Figure \ref{fig2}(b), the mode activates at a positive critical field $B_{c1} \approx +2.5$ T. Once activated, $M^*$ persists as the magnetic field is reduced from +7 T down to zero. The integrated Raman scattering of the $M^*$ Raman excitation remains nearly constant across zero-field until a negative critical field $B_{c2} \approx -3.5$ T is reached, where the mode is suppressed. This asymmetry reveals a non-reciprocal response of the field-induced excitation. 

To determine how this new mode correlates with the antiferromagnetic order of the Ru$^{4+}$ spins, we investigate the temperature dependence of the $B_{1g}$ Raman spectra at a fixed field $B = +4$ T above $B_{c1}$. The $B_{1g}$ spectra are presented in Figure \ref{fig2}(c), while the integrated Raman scattering of the $M^*$ is reported in Figure \ref{fig2}(d). The spectral weight of $M^*$ closely tracks the temperature dependence of the conventional in-plane magnons, as shown in the right panels of Figure \ref{fig2}(c). Upon reaching the N\'eel temperature ($\TN$), the $M^*$ mode completely disappears, remaining absent throughout the paramagnetic phase. Note that the experimentally observed $\TN$ in our Raman measurements is lower than the nominal bulk value (110 K) due to local heating induced by the CW laser probe. 

The temperature dependence of $M^*$ demonstrates that the field-induced excitation is linked to the underlying long-range antiferromagnetic order. Moreover, the abrupt activation of $M^*$ at finite critical fields further supports its magnetic origin over a continuous field-induced structural distortion, such as a zone-folded phonon.

\section{Hysteretic evolution of the Fano anomaly of the $H$ phonon}

\begin{figure*}[t]
\centering
\includegraphics[width=0.63\textwidth]{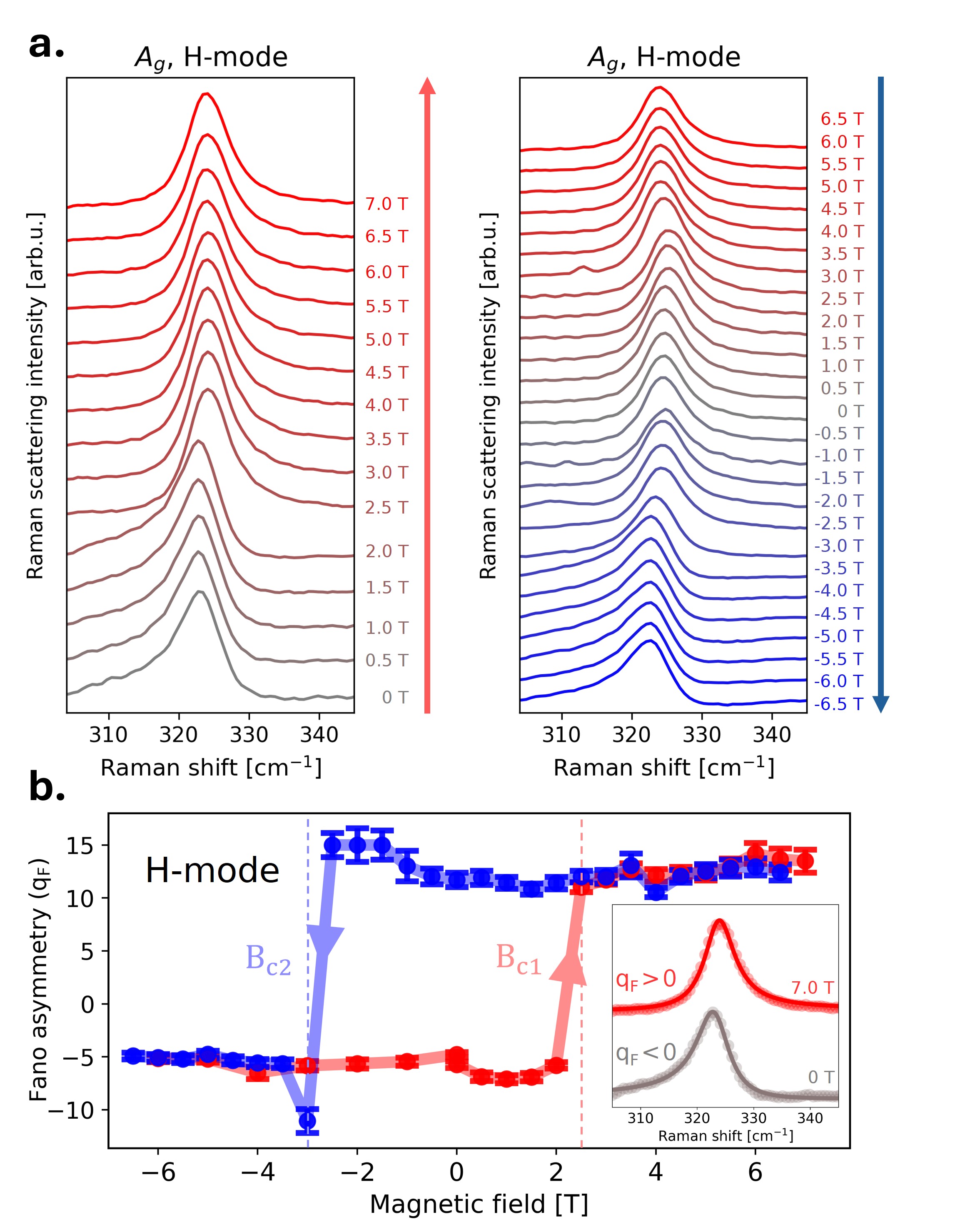}
\caption{\textbf{Evolution of the Fano anomaly of the $H$ mode in magnetic field.}
\textbf{(a)} Raman scattering response of the $A_g$-symmetry $H$ mode (324 cm$^{-1}$) of Ca$_2$RuO$_4$ at 1.6 K as a function of the magnetic field, measured during increasing-field (left) and decreasing-field (right) sweeps. \textbf{(b)} Magnetic-field dependence of the Fano asymmetry parameter $q_{\mathrm{F}}$, obtained from fits of the $H$ mode with a Fano profile. Representative fits below ($B = 0$ T) and above ($B = +7$ T) the critical field are shown in the inset. The $H$ mode shows a non-reciprocal hysteretic sign change of the asymmetry parameter $q_{\mathrm{F}}$, consistent with the non-reciprocal hysteresis of the $M^*$ mode observed in the $B_{1g}$ channel.}
\label{fig:fig3}
\end{figure*}

Given the evidence of strong magneto-elastic coupling in the AF phase of Ca$_2$RuO$_4$~\cite{braden1998, rho2005, lee2019}, we investigate how the lattice phonons are affected by the onset of the $M^*$ mode through the magnetic-field cycle. Specifically, we focus our attention on the $H$ phonon mode, which exhibits a pronounced lineshape asymmetry even in the zero-field state, providing a direct signature of its strong magneto-elastic coupling with the magnetic Higgs excitation. 

Figure \ref{fig:fig3}(a) illustrates the Raman response of the $H$ mode during a complete magnetic-field cycle, showing the spectra for both the increasing and decreasing field sweeps in the range from $-6.5$ T to $+7$ T. The lineshape of the $H$ mode undergoes a hysteretic reconstruction, visibly changing its asymmetry.

To quantify this behavior, we fit the $H$ phonon profile using the Fano lineshape function~\cite{fano1961, klein1975}. This profile is characteristically employed for Raman modes where a discrete excitation, such as a structural phonon, hybridizes with a continuum of background excitations~\cite{lemmens2003}. In Ca$_{2}$RuO$_{4}$, the resulting lineshape asymmetry serves as a sensitive probe to map the underlying magneto-elastic coupling between this specific lattice mode and the magnetic Higgs excitation, which is peaked at the $H$ phonon frequency~\cite{souliou2017}. The Fano fitting function is defined as:
\begin{equation}
I(\omega) = I_0 \frac{(q_{\mathrm{F}} + \epsilon)^2}{1 + \epsilon^2},
\label{eq:fano}
\end{equation}
where $I_0$ is the intensity of the uncoupled phonon mode, $q_{\mathrm{F}}$ is the phonon asymmetry parameter, which quantifies the strength of the coupling with the continuum excitations, and $\epsilon = (\omega - \omega_0)/\Gamma$ is the reduced frequency normalized to the mode linewidth $\Gamma$.

As shown in Figure \ref{fig:fig3}(b), the field dependence of $q_{\mathrm{F}}$ (extracted from the individual Fano fits, see Figure~\ref{fig:S5} of the SM) traces an asymmetric, non-reciprocal hysteresis loop. Crucially, $q_{\mathrm{F}}$ exhibits a hysteretic switch between negative and positive values at the critical fields $B_{c1} \approx +2.5$ T and $B_{c2} \approx -3.0$ T, following the emergence of the magnetic mode $M^*$ in the $B_{1g}$ channel (Figure \ref{fig2}(b)). This concurrent behavior directly reflects a field-induced modification of the magneto-elastic coupling with the Higgs mode. As suggested by the differential Raman spectra detailed in the SM (Figure \ref{fig:S4}), the external magnetic field drives a spectral redistribution in the frequency region of the underlying Higgs mode. Through magneto-elastic coupling, this spectral redistribution is directly reflected onto the hysteretic modification of the Fano asymmetry parameter $q_{\mathrm{F}}$ of the $H$ phonon. Moreover, we note that $q_{\mathrm{F}}$ undergoes a sign change across the magnetic-field cycle. Since the magnetic Higgs mode is known to exhibit a broad, multi-component structure in frequency~\cite{souliou2017}, this sign reversal points towards a scenario in which the centroid of the magnetic Higgs continuum shifts across the localized $H$ phonon frequency, rather than a change in the sign of the Higgs-phonon coupling. To verify that this mechanism is linked to the magneto-elastic coupling with the Higgs mode, we track another $A_g$ phonon at 395 cm$^{-1}$, which is out-of-resonance with the peak of the Higgs excitation. As detailed in Figure \ref{fig:S_Ag_phon_fits} of the SM, the 395 cm$^{-1}$ $A_g$ phonon exhibits no significant modification of its Fano parameter across the same magnetic-field loop. 

The behavior of the Fano anomaly of the $H$ phonon hence provides a structural fingerprint of both the observed magnetic-field transition and the underlying modification of the Higgs mode spectral weight.

\section{Discussion}

\begin{figure*}[t!]
    \centering
\includegraphics[width=0.93\textwidth]{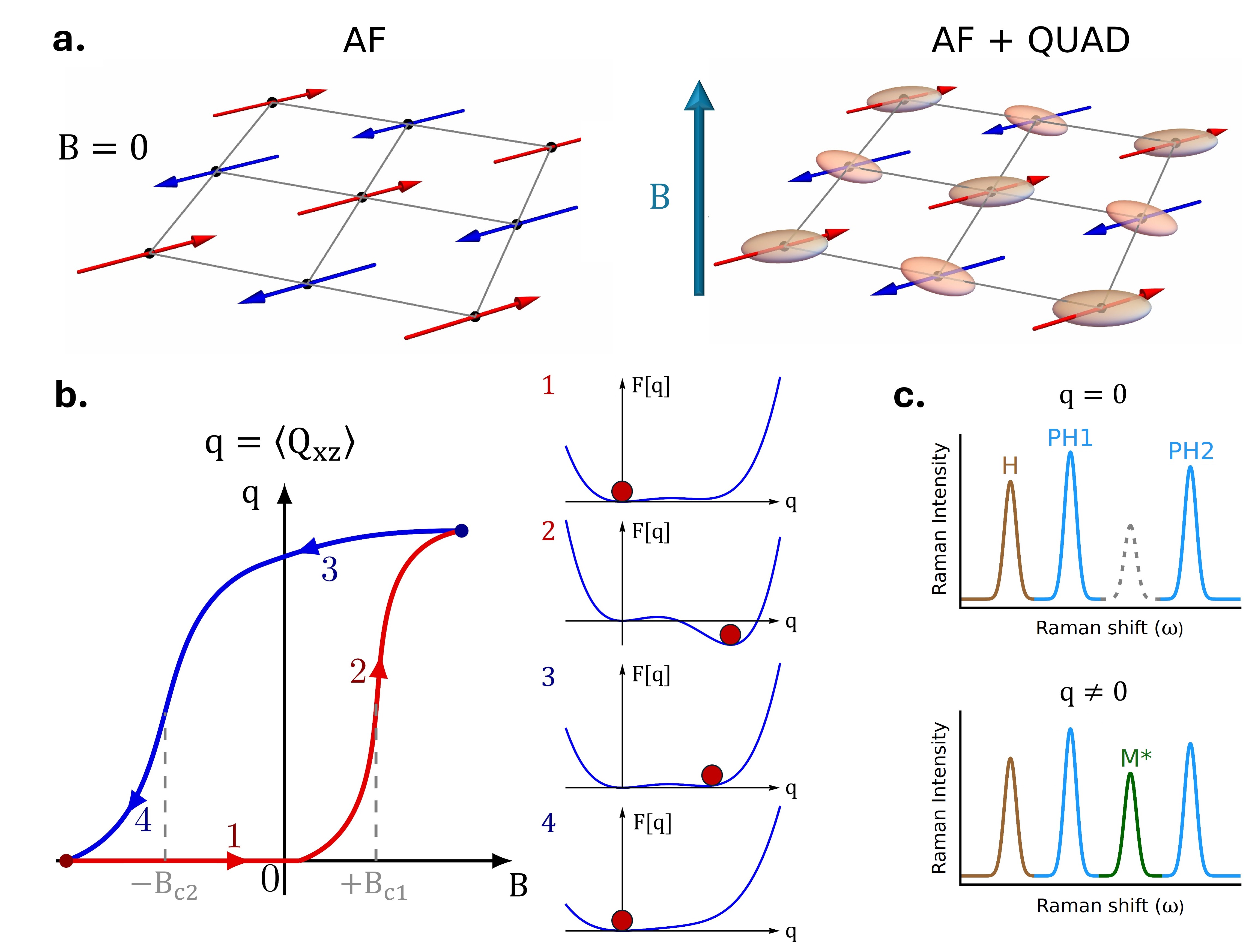}

\caption{\textbf{Mechanism for the induced quadrupolar order and activation of the $M^*$ Raman mode in $c$-axis magnetic field.} \textbf{(a)} Schematic representation of the antiferromagnetic (AF) phase (left) and the coexisting antiferromagnetic–quadrupolar (AF+QUAD) phase activated by the $c$-magnetic field (right). \textbf{(b)} Hysteresis cycle of the quadrupolar order parameter $q \equiv \langle Q_{xz} \rangle$ as a function of the applied magnetic field. For the four representative points along the hysteresis cycle, the corresponding free-energy profiles as functions of $q$ are shown on the right, with the red markers indicating the system residing in a stable or metastable minimum. \textbf{(c)} Sketch of the predicted Raman spectrum within the energy window of interest. When $q=0$ (top), three peaks appear: the $H$ phonon (brown curve) and two phonon modes ($PH1$ and $PH2$, light blue curves), while the mode $M^*$ (dotted gray curve) is absent. For $q\neq0$ (bottom), a finite coupling between the Higgs mode and the $H$ phonon mode is switched on, leading to the activation of the additional mode $M^*$ (green curve).}
  \label{fig4}
\end{figure*}

Our Raman data provide evidence for an anomalous first-order magnetic phase transition driven by an out-of-plane magnetic field. The most striking signature of this transition is the emergence of the $M^*$ mode. As detailed in the SM (Section \ref{section:S2}), applying a magnetic field along the $c$-axis lowers the symmetry to the magnetic point group $m^\prime m^\prime m$. According to group theory, this causes the $A_g$ and $B_{1g}$ irreducible representations to form the corepresentation $DA_g$, meaning these channels mix and inherit each other's Raman activity. However, since DFT calculations rule out a structural Raman phonon at the frequency of $M^*$, its observation implies a further field-induced symmetry breaking. Crucially, the invariance of the in-plane magnons demonstrates that the primary antiferromagnetic dipolar order remains intact under the applied magnetic field. This additional symmetry breaking therefore hints towards a non-dipolar mechanism, consistent with the emergence of a quadrupolar sector of the magnetic moments, as schematically illustrated in Figure~\ref{fig4}(a).

In a spin-1 quantum magnet, longitudinal fluctuations of the ordered magnetic moment necessarily involve changes in the local quadrupolar degrees of freedom. As a result, the Higgs mode ~\cite{jain2017, souliou2017}, the collective amplitude mode of the magnetic order parameter, has an intrinsic quadrupolar character and can naturally couple to quadrupolar excitations.

This physical picture is corroborated by two main experimental observations. First, the emergent $M^*$ mode acts as a Higgs-coupled mode, originating from the hybridization between the Higgs mode and a structural phonon. This is evidenced by the fact that $M^*$ completely disappears above the N\'eel temperature ($\TN$), where the antiferromagnetic order and its associated Higgs mode are no longer present. 

Second, the magnetic field drives a hysteretic reconstruction of the Fano asymmetry parameter ($q_{\mathrm{F}}$) of the $H$ phonon. Since this asymmetric lineshape originates from the coupling between the structural mode and the underlying Higgs continuum, its modification provides direct evidence that the external field alters the magneto-elastic coupling. Essentially, the magnetic field reshuffles the spectral weight of the Higgs excitation (see SM, Figure \ref{fig:S4}), thereby changing how it couples to the lattice. 

To rationalize the emergence of this quadrupolar sector, one must consider the local electronic structure of Ca$_2$RuO$_4$. Due to the competition between spin-orbit coupling and electron-lattice interactions, the local physics of the Ru$^{4+}$ ions can be mapped onto a low-energy manifold with an effective local pseudospin-$1$ moment~\cite{khaliullin2013}. Unlike $S=1/2$ systems, an $S=1$ local Hilbert space naturally accommodates not only magnetic dipole moments but also rank-2 spin quadrupoles, $Q^{\alpha\beta}=S^\alpha S^\beta+S^\beta S^\alpha-(2/3)\delta_{\alpha\beta}$, thereby allowing phases beyond standard N\'eel order. Choosing a coordinate system in which $x$ lies along the in-plane staggered magnetization ($n_x$) and $z$ is aligned with both the external field $\mathbf{B}=B\mathbf{z}$ and the canted component of the staggered moment ($n_z$), the relevant degree of freedom reduces to the $xz$ quadrupolar component, hereafter denoted as $q \equiv Q_{xz}$.

The magnetic-field-induced transition, together with the non-reciprocal hysteretic behavior observed in the amplitude of the $M^*$ mode and in the Fano anomaly of the $H$ phonon, can be rationalized within a phenomenological scenario in which the quadrupolar order parameter $q$ is already close to an instability at zero external field, featuring a global minimum at $q=0$ and a metastable minimum at $q\neq 0$, separated by an energy barrier. An external magnetic field applied along the out-of-plane direction then couples simultaneously to the N\'eel vector and $q$, thus softening or hardening the stiffness of quadrupolar fluctuations and favoring either the $q=0$ or the $q\neq 0$ minimum depending on the orientation of $B$. Because of the energy barrier between local minima, the system can remain trapped in a local minimum depending on its history, leading naturally to hysteresis and, in the presence of odd-in-$B$ terms (i.e., if the evolution under positive and negative magnetic field is not equivalent), to non-reciprocal hysteretic behavior. This physical scenario is illustrated in Figure\ \ref{fig4}(b).

Finally, the quadrupolar order parameter couples linearly to spin and lattice strain via magneto-elastic interactions, such that a finite quadrupolar amplitude renormalizes the spin–phonon coupling and induces hybridization between the $H$ phonon mode and the Higgs mode. This mixing gives rise to the additional collective excitation, the $M^*$ mode, closely tied to the quadrupolar sector, which becomes active upon the onset of quadrupolar order, providing an indirect experimental signature of the non-trivial quadrupolar phase. This mechanism is sketched in Figure~\ref{fig4}(c), where $PH1$ and $PH2$ denote the two $B_{1g}$ phonon modes at 395 cm$^{-1}$ and $460$ cm$^{-1}$, respectively.

In the framework of Landau phase transitions, the minimal effective free energy (per unit volume) describing the quadrupolar instability in the absence of a magnetic field and neglecting dipole-quadrupole interactions reads
%
\begin{equation}
\mathcal{F}_q=
\frac{r}{2}q^2-
\frac{s}{3}q^3+
\frac{u}{4}q^4.
\label{Fq}
\end{equation}
The stiffness $r$ and the coefficients $s$ and $u$ are all positive. In particular, for $s^2>4ru$, $\mathcal{F}_q$ exhibits two minima at $q=0$ and $q=q^*$, with $q^*\equiv\left( s+\sqrt{s^2-4ru} \right)/(2u)$.

When dipole–quadrupole interactions are taken into account, one must first consider the bare free energy of the staggered magnetic moment $n\equiv\sqrt{n_x^2+n_z^2}$, which reads $\mathcal{F}_n=-(r_n/2)n^2+(u_n/4)n^4$, with $r_n,u_n>0$. $\mathcal{F}_n$ describes AF ordering with staggered moment $\overline{n}\equiv\sqrt{r_n/u_n}$. Then, at lowest order in $n_x$, $n_z$ and $q$, symmetry allows a trilinear coupling $\lambda  n_x n_z q$. Since in Ca$_2$RuO$_4$ the component $n_z$ is weak but finite already at $B=0$~\cite{porter2018}, the quadrupolar sector experiences a small symmetry-breaking field $\lambda \theta n^2$, where $\theta$ is the small canting angle present even in the absence of an applied field. This feature naturally leads to an asymmetric hysteretic field when $B\neq 0$. In addition, the magnetic field couples to the quadrupolar sector through a term $\mu B n_x q$ at leading order, so that the total coupling reads 
\begin{equation}
\mathcal{F}_{nq}=
\lambda\theta n^2 q+
\mu B n q.
\label{Fnq}
\end{equation}

The first effect of $\mathcal{F}_{nq}$ is to slightly renormalize the staggered magnetization. Therefore, $\partial_n\left(\mathcal{F}_n + \mathcal{F}_{nq}\right)=0$ can be solved perturbatively for $n$, as detailed in the SM. Substituting this solution back into the total free energy yields an effective free energy in terms of $q$ only, in which the quadrupolar stiffness is renormalized as $\tilde{r}=r-\kappa B$, with $\kappa\equiv 2\lambda \mu\theta\sqrt{r_n u_n}$. That is, the inclusion of $B$, along with the trilinear coupling between the staggered components and the quadrupolar order, leads to an additional term $-(\kappa/2)B q^2$ in $\mathcal{F}_q$. This term is responsible for the softening or hardening (depending on the sign of $B$) of the quadrupolar stiffness, thereby driving first-order transitions between $q=0$ and $q=\tilde{q}^*$, with $\tilde{q}^*\equiv\left( s+\sqrt{s^2-4\tilde{r}u} \right)/(2u)$. Moreover, this term is odd in $B$ and therefore accounts for the non-reciprocal hysteretic behavior. \\
\indent At zero magnetic field, the system resides in the $q=0$ minimum. Upon applying a positive magnetic field $B>0$, the transition to $q=\tilde{q}^*$ (i.e., from point 1 to point 2 in the hysteresis diagram of Figure\ \ref{fig4}(b)) occurs at the critical field $B=B_{c1}\equiv r/\kappa$, where the effective stiffness $\tilde{r}$ vanishes. Upon decreasing the magnetic field from $B_{c1}$ towards negative values, the system remains trapped in the metastable state $q=\tilde{q}^*$ until the reverse transition to $q=0$ (from point 3 to point 4 in Figure\ \ref{fig4}(b)) occurs at $B=-B_{c2}$, with $B_{c2}\equiv s^2/(4 u\kappa)-r/\kappa$ (see SM, Section \ref{sec:free_energy}). Since $B_{c1}\neq B_{c2}$, the predicted hysteresis is non-reciprocal with respect to the zero-magnetic-field AF state. A quantitative estimate of the two critical fields is beyond the scope of this paper. As a final remark, we note that for $s^2>8 ru$, one has $B_{c2}>B_{c1}$, in agreement with the experimentally observed critical fields.

Finally, we highlight that the proposed scenario for the activation of the quadrupolar phase relies on a single-domain framework. Indeed, the non-reciprocal hysteresis of the quadrupolar order parameter emerges intrinsically within a single-domain picture, avoiding the need to invoke domain-wall dynamics. Future investigations utilizing spatially resolved techniques will be necessary to directly probe the local magnetism.

\section{Conclusions}

Using polarization-resolved Raman spectroscopy, we report an anomalous magnetic phase transition in the antiferromagnetic phase of Ca$_2$RuO$_4$ driven by an out-of-plane magnetic field. The transition is anomalous in the sense that the magnetic field does not alter the in-plane magnons, demonstrating that the standard dipolar magnetic moments in the $ab$ plane are preserved. Instead, the field selectively reconstructs the low-energy Raman spectrum, activating a new magnetic mode ($M^*$) and driving a sign reversal of the Fano anomaly for a lattice mode ($H$ phonon). Crucially, both spectral features exhibit a non-reciprocal hysteresis in magnetic field. We rationalize these observations as the field-induced activation of the quadrupolar sector of the magnetic moments, which can alter the magneto-elastic coupling and trigger a Higgs-phonon hybridization. This is captured by a phenomenological Ginzburg-Landau framework, where the quadrupolar order parameter follows the observed non-reciprocal hysteresis of the Raman modes. This evidence can open the way to selectively control new interacting phases by exploiting magnetic-Higgs-like modes and quadrupolar degrees of freedom in spin-orbit Mott insulators.

\begin{acknowledgments}

The authors acknowledge the support of the Gordon and Betty Moore Foundation through the grant CENTQC (no. GBMF12213). This research was partly funded by the Deutsche Forschungsgemeinschaft (DFG, German Research Foundation)-TRR 360-492547816.

\end{acknowledgments}

%
%

\clearpage
\onecolumngrid

\setcounter{section}{0}
\setcounter{subsection}{0}
\setcounter{equation}{0}
\setcounter{figure}{0}
\setcounter{table}{0}

\renewcommand{\thesection}{S\arabic{section}}
\renewcommand{\thesubsection}{S\arabic{section}.\arabic{subsection}}
\renewcommand{\theequation}{S\arabic{equation}}
\renewcommand{\thefigure}{S\arabic{figure}}
\renewcommand{\thetable}{S\arabic{table}}

\renewcommand{\theHsection}{S\arabic{section}}
\renewcommand{\theHfigure}{S\arabic{figure}}
\renewcommand{\theHequation}{S\arabic{equation}}
\renewcommand{\theHtable}{S\arabic{table}}

\makeatletter
\renewcommand{\section}{\@startsection{section}{1}{\z@}%
  {-3.5ex \@plus -1ex \@minus -.2ex}%
  {2.3ex \@plus.2ex}%
  {\normalfont\bfseries}}
\makeatother

\begin{center}
{\Large\bfseries Supplementary Material for}\\[0.2cm]
{\Large\bfseries Raman signatures of a non-reciprocal magnetic phase transition in Ca$_2$RuO$_4$}\\[0.35cm]

Giacomo Jarc$^{1,*}$, Giovanni Tartaglia$^{2,*}$, Francesco Gabriele$^{3}$,
Filomena Forte$^{3}$,  Anita Guarino$^{3}$, Angela Montanaro$^{1}$, Enrico Maria Rigoni$^{1}$, Nitesh Khatiwada$^{1}$, Costanza Lincetto$^{1}$, Gabriele Bartolini$^{1}$, Antonio Mastropasqua$^{1}$, Shahla Yasmin Mathengattil$^{2, 4}$ , Marco Malvestuto$^{4, 5}$, Muhammad Waqee Ur Rehman$^{3}$, Rosalba Fittipaldi$^{3}$, Joachim Deisenhofer$^{6}$, Alexander A. Tsirlin$^{7}$, Antonio Vecchione$^{3}$, Mario Cuoco$^{3}$, Daniele Fausti$^{1,\dagger}$

\vspace{0.25cm}

{\itshape
$^{1}$Department of Physics, University of Erlangen-Nürnberg, Erlangen, Germany \\
$^{2}$Department of Physics, University of Trieste, Trieste, Italy \\
$^{3}$CNR-SPIN, University of Salerno, Fisciano, Salerno, Italy \\
$^{4}$Elettra Sincrotrone Trieste, Trieste, Italy \\
$^{5}$CNR-Istituto Officina dei Materiali (IOM), Trieste, Italy}\\
$^{6}$Experimental Physics V, Center for Electronic
Correlations and Magnetism, Institute for Physics, University of Augsburg, D-86159 Augsburg, Germany\\
$^{7}$Felix Bloch Institute for Solid-State Physics, Leipzig University, 04103 Leipzig, Germany

\vspace{0.5cm}

{\small
$^{*}$These authors contributed equally to this work.\\
$^{\dagger}$Corresponding author: \href{mailto:daniele.fausti@fau.de}{daniele.fausti@fau.de}
}
\end{center}

\clearpage

\section{\protect\NoCaseChange{Temperature-dependent Raman scattering from Ca$_2$RuO$_4$ across the antiferromagnetic transition}}

\begin{figure}[hbt!]
\centering
\includegraphics[width=\textwidth]{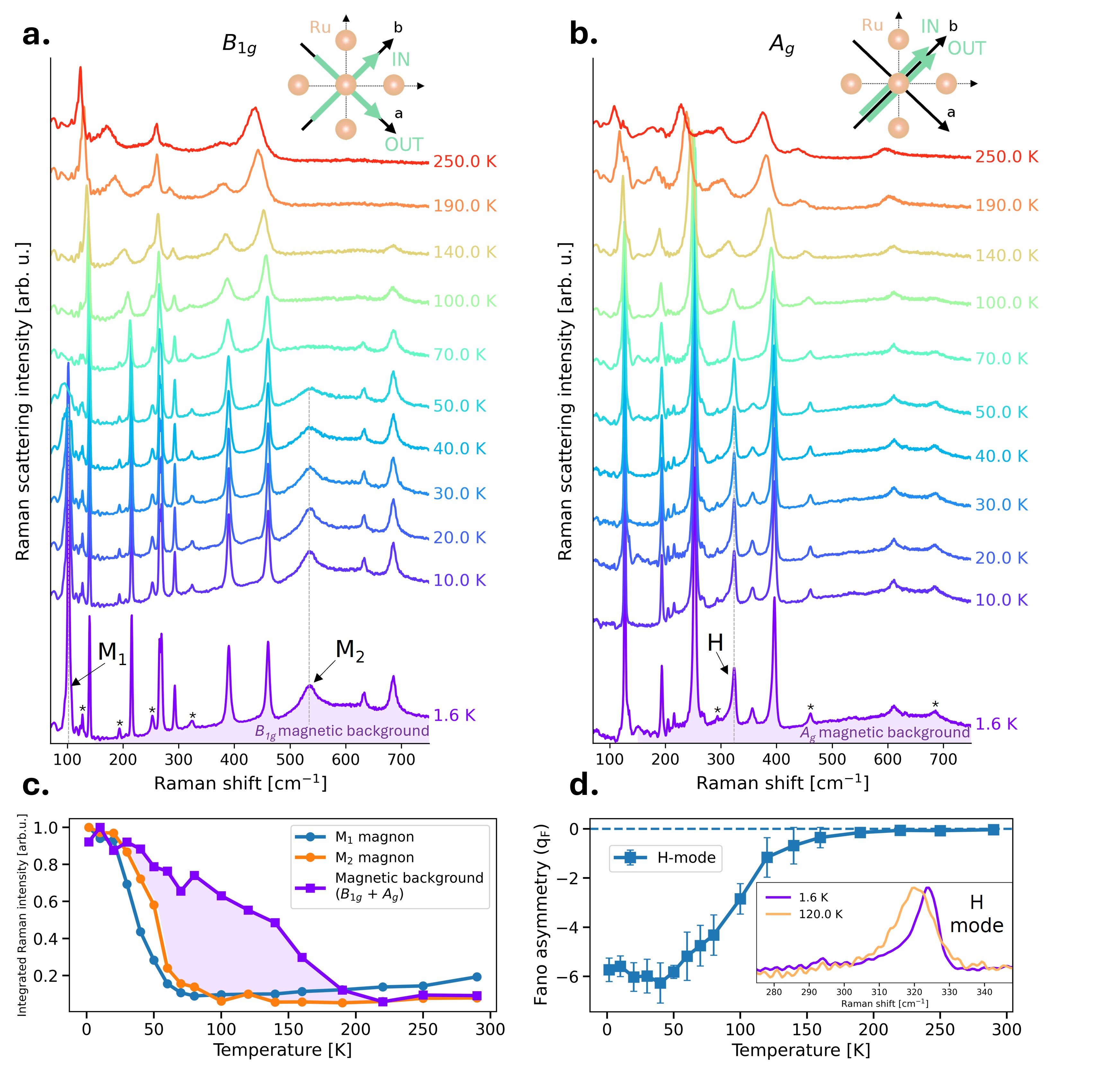}
\caption{\textbf{Temperature-dependent Raman scattering from Ca$_2$RuO$_4$ across the antiferromagnetic transition.}
Raman scattering from Ca$_2$RuO$_4$ as a function of temperature measured in the $B_{1g}$ \textbf{(a)} and $A_g$ \textbf{(b)} polarization geometries. The insets indicate the incident (IN) and scattered (OUT) polarizations relative to the Ru-O bonds for the $B_{1g}$ (a) and $A_g$ (b) configurations. The spectra are vertically shifted for clarity. The $M_1$ and $M_2$ magnon modes of the in-plane antiferromagnetic state are highlighted, together with the broad magnetic background visible in both $B_{1g}$ and $A_g$ geometries. In $A_g$, the $H$ phonon mode is highlighted. Asterisks mark modes from the complementary channel. \textbf{(c)} Temperature dependence of the integrated Raman spectral weight of the $M_1$ and $M_2$ magnons and of the total magnetic background, obtained as the sum of the $B_{1g}$ and $A_g$ scattering. All the integrals are normalized to the maximum. \textbf{(d)} Temperature dependence of the fitted Fano-asymmetry parameter $q_{\mathrm{F}}$ of the $H$ phonon mode; the inset compares the $H$-mode lineshape below and above $T_N$, showing a stronger asymmetry in the AF-ordered state.}
\label{S1}
\end{figure}

In Figure \ref{S1}, we detail the temperature dependence of the Raman spectra of Ca$_2$RuO$_4$ at zero magnetic field in both high-symmetry configurations. Figure \ref{S1}(a) displays the raw spectra in the $B_{1g}$ scattering geometry as a function of temperature. This channel identifies the $M_1$ single-magnon and the $M_2$ two-magnon excitations, which are superimposed on a broad continuum of magnetic scattering. The asterisks mark phonon modes from the complementary polarization channel. These residual modes are symmetry-allowed in the antiferromagnetic phase according to the corepresentation analysis presented in Section \ref{section:S2}.

Figure \ref{S1}(b) presents the temperature-dependent spectra in the $A_g$ geometry, featuring the $H$ phonon mode. At low temperatures, this phonon exhibits an asymmetric Fano-like lineshape due to its coupling with the underlying magnetic continuum. It is crucial to note that the magnetic background in this $A_g$ channel is composite. Specifically, it consists of the magnetic Higgs mode, which is strictly confined to the $A_g$ symmetry, peaks around the frequency of the $H$ phonon (324 cm$^{-1}$), and vanishes at the N\'eel temperature ($T_{\mathrm{N}}$), superimposed on a broader scattering continuum. Crucially, while the Higgs excitation is symmetry-restricted to the $A_g$ channel, the scattering from short-range antiferromagnetic correlations is isotropic and manifests in both the $A_g$ and $B_{1g}$ geometries. While the Higgs contribution melts together with the long-range order at $T_{\mathrm{N}}$, these short-range fluctuations persist well into the paramagnetic phase before eventually being suppressed at higher temperatures ($\sim$ 200 K).

To capture the full temperature evolution of these shared short-range fluctuations, Figure \ref{S1}(c) tracks the temperature dependence of the total integrated magnetic background ($A_g + B_{1g}$), alongside the spectral weight of the $M_1$ and $M_2$ magnons across the AF transition. The intensity of the $M_1$ and $M_2$ magnons, which probe the long-range antiferromagnetic order, collapses at $T_{\mathrm{N}}$. As noted in the main text, the experimentally observed $T_{\mathrm{N}}$ is lower than the nominal value due to local heating from the CW probe ($P_{\mathrm{CW}} = 3$ mW, $\sim$10 $\mu$m focus size).

Finally, Figure \ref{S1}(d) tracks the temperature evolution of the Fano asymmetry parameter $q_{\mathrm{F}}$ of the $H$ mode. This lineshape asymmetry is intrinsically linked to the Higgs amplitude mode of the spin-orbit condensate; consequently, $q_{\mathrm{F}}$ decreases significantly as the system approaches $T_{\mathrm{N}}$. Crucially, the asymmetry parameter does not abruptly drop to zero immediately above $T_{\mathrm{N}}$. Because the sharp $H$ excitation sits directly on top of the magnetic scattering continuum, and since the short-range antiferromagnetic fluctuations survive above $T_{\mathrm{N}}$ in both channels (as evidenced by the high-temperature total background in Figure \ref{S1}(c)), they continue to provide a scattering continuum for the phonon. This prevents a recovery of a symmetric lineshape ($q_{\mathrm{F}} \rightarrow 0$) until higher temperatures are reached ($\sim$ 160 K).

\clearpage
\newpage

\section{\protect\NoCaseChange{Symmetry analysis and optical phonons in Ca$_2$RuO$_4$}}
\label{section:S2}
In order to clarify the assignment of the observed Raman modes, we shortly discuss the expected phonon selection rules for the paramagnetic and antiferromagnetic phase and in the presence of a magnetic field along the crystallographic $c$-axis:
The paramagnetic space group of Ca$_2$RuO$_4$ is orthorhombic  $Pbca$ (No. 61) and its point group is $G=mmm$ \cite{Porter:2018}
\begin{equation}
    \mathbf{G}=\{E, C_{2x}, C_{2y},C_{2z}, I,m_x,m_y,m_z\}
\end{equation}

The corresponding irreducible representation for Ca$_2$RuO$_4$ with four formular units in the primitive unit cell yields 84 normal modes at the $\Gamma$-point \cite{Rho:2005}
\begin{align}\label{Eq:irreps}
\Gamma &=  11B_{1u}(z) +11B_{2u}(y) +11B_{3u}(x) &&\text{(IR)}\nonumber \\ 
&+  9A_g (x^2,y^2,z^2) + 9B_{1g} (xy)  &&\text{(Raman)}\nonumber \\ 
&+ 9B_{2g} (xz) +9B_{3g}(yz) &&\text{(Raman)}\nonumber \\
&+ B_{1u}+ B_{2u}+ B_{3u}  &&\text{(acoustic)}\nonumber \\ 
&+ 12A_u  &&\text{(silent)}     
\end{align}

Below the N\'eel temperature, Ca$_2$RuO$_4$ exhibits a $k=0$ antiferromagnetic order and the magnetic unit cell is identical to the paramagnetic one \cite{Porter:2018}, i.e. no appearance of new modes due to Brillouin-zone folding is expected. Moreover, the magnetic space group remains $Pbca$ (No.~61.433) and the magnetic point group remains $mmm$, i.e. the selection rules for the optical phonons remain unchanged by the magnetic ordering in agreement with our data and previous Raman studies \cite{Rho:2005}. Hence, we conclude that any additional mode appearing in the magnetically ordered phase is of magnetic origin.

To determine possible changes in selection rules in the presence of a magnetic field applied along the $c$-axis, one additionally has to ensure that the external field is invariant under all symmetry operations of the corresponding point group. Both in the paramagnetic and antiferromagnetic phase this leads to the magnetic point group $M=m'm'm$
\begin{equation}
    \mathbf{M}=\{E,I, C_{2z}, m_z, C^\prime_{2x},C^\prime_{2y}, m_x^\prime,m_y^\prime\},
\end{equation}
where the two-fold rotations around the $x$- and $y$-axis and the corresponding mirror planes have to be combined with the anti-unitary time-reversal operation $1'$ \cite{Bradley:2009}.

The irreducible corepresentations of the magnetic point group can be determined by the procedure of Anastassakis and Burstein \cite{Anastassakis:1972} and result in
\begin{align}\label{Eq:irreps}
D\Gamma &=  23DA_{u}(z) +22DB_{u}(x,y) &&\text{(IR)}\nonumber \\ 
&+  18DA_g (x^2,y^2,z^2,xy) + 18DB_{g} (xz,yz) &&\text{(Raman)}\nonumber \\ 
&+ DA_{u}+ 2DB_{u}  &&\text{(acoustic)}\nonumber \\    
\end{align}
The explicit reduction of $mmm$ with respect to the unitary halving subgroup $H=2_z/m_z$ is given in Table~\ref{tab:Coreps_cro}:\\
The 9$A_g$ and 9$B_{1g}$ modes now form the corepresentation $DA_g$ with 18 modes, which all inherit the Raman activity of both the  $A_g$ and $B_{1g}$ channels. Similarly, the $B_{2g}$ and $B_{3g}$ now form the corepresentation $DB_g$. The odd irreducible representations undergo the same changes, where it is noteworthy to mention that the previously silent 12$A_u$ modes now become infrared active for light polarization along the $z$-axis as a part of the $DA_u$ corepresentation. It is important to note that although the corepresentation analysis predicts the mixing of the $A_g$ and $B_{1g}$ channels, it cannot give the matrix elements that determine the Raman scattering intensities in the symmetry-reduced state.

\begin{table}[H]
    \centering
    \begin{tabular}{l|l|c|c}
      $\mathbf{g}=\mathbf{G} \oplus\{E+\mathcal{R}\}$   & $\mathbf{G}$ & $\mathbf{H}$ &  $\mathbf{M}=\mathbf{H}+\mathcal{R}(\mathbf{G}-\mathbf{H})$ \\
         & $mmm$ & $2/m$ &  $m^\prime m^\prime m$ \\\hline
    DA$_g$  &  A$_g$ & A$_g$ & DA$_g$\\
    DB$_{1g}$  &  B$_{1g}$ & A$_g$ & DA$_g$\\
    DB$_{2g}$  & B$_{2g}$ & B$_{g}$ &   DB$_{g}$\\
    DB$_{3g}$  &  B$_{3g}$ & B$_{g}$ &   DB$_{g}$\\
    DA$_u$ &  A$_u$ & A$_u$ & DA$_u$\\
    DB$_{1u}$  &  B$_{1u}$ &  A$_u$ & DA$_u$\\
    DB$_{2u}$  &  B$_{2u}$ & B$_{u}$ & DB$_{u}$\\
    DB$_{3u}$  &  B$_{3u}$ & B$_{u}$ & DB$_{u}$\\
    \end{tabular}
    \caption{Reduction of the irreducible representations of the nonmagnetic point group $\mathbf{G}=mmm$ with respect to the unitary halving subgroup $\mathbf{H}=2/m$ to determine the corepresentations of the magnetic point group $\mathbf{M}=m^\prime m^\prime m$.}
    \label{tab:Coreps_cro}
\end{table}

We conclude that in our Raman configurations additional modes, which appear only in the presence of an applied magnetic field such as mode $M^*$ are either of magnetic origin or indicate a further symmetry reduction of the lattice via magneto-elastic effects.

In order to clearly assign the observed Raman modes, we calculate the eigenfrequencies of $A_g$ and $B_{1g}$ and compare them to the experimental values in Table~\ref{tab:eigenfrequencies}. Note that the agreement of experimental and calculated eigenfrequency can be considered good for the $B_{1g}$ modes and the first five $A_g$ modes with eigenfrequencies below the H mode. For the higher-lying $A_g$ modes the deviations are considerable and might indicate the influence of fluctuation and hybridization effects, which are not considered in the calculations.

\begin{table}[H]
\centering
\begin{tabular}{cc|cc}
\hline
\multicolumn{2}{c|}{$A_g$} & \multicolumn{2}{c}{$B_{1g}$} \\
\makebox[1.5cm]{Exp.} & \makebox[1.5cm]{Calc.} & \makebox[1.5cm]{Exp.} & \makebox[1.5cm]{Calc.} \\
\hline
127 & 124.1 & 140 & 135.2 \\
198 & 187 & 215 & 227.6 \\
204 & 209.2 & 266 & 256 \\
253 & 256.5 & 269 & 263.6 \\
304 & 306.9 & 293 & 287.6 \\
\,\,\,324\textsuperscript{\emph{a}} & 333.8 & 390 & 381.7 \\
356 & 409.5 & 461 & 468.8 \\
395 & 610.3 & \,\,\,534\textsuperscript{\emph{b}} & 531.8 \\
607 & 638.2 & 686 & 667.9 \\
\hline
\multicolumn{4}{l}{\footnotesize \textsuperscript{\emph{a}} H mode.} \\
\multicolumn{4}{l}{\footnotesize \textsuperscript{\emph{b}} Superimposed with the 2-magnon mode.}
\end{tabular}
\caption{Experimental Raman excitation frequencies in Ca$_2$RuO$_4$ (in cm$^{-1}$) measured in  $z(\parallel)\bar{z}$ and  $z(\perp)\bar{z}$ configurations in the magnetically ordered phase at 1.6~K. Mode assignment is made by comparison with calculated phonon eigenfrequencies obtained from \textit{ab initio} calculations for the 9 expected $A_g(i)$ modes $i=1,\dots,9$ and to the nine expected $B_{1g}(j)$ modes $j=1,\dots,9$ .}
\label{tab:eigenfrequencies}
\end{table}

Density-functional theory (DFT) band structure calculations were performed in the VASP code~\cite{vasp1,vasp2} using the Perdew-Burke-Ernzerhof version of the exchange-correlation potential~\cite{pbe96} and the mean-field DFT+$U$ correction for the correlated Ru $4d$ states with $U_d=3$ eV and $J_d=0.5$ eV~\cite{bader2023}. Antiferromagnetic spin configuration was chosen. Experimental low-temperature lattice parameters from Ref.~\cite{braden1998} were used, and the atomic positions were fully relaxed prior to phonon calculations. Phonon frequencies were obtained using frozen atomic displacements. The $8\times 8\times 4$ $k$-mesh was used in all calculations.

\clearpage
\newpage

\section{\protect\NoCaseChange{Evolution of the Raman scattering from the in-plane magnons in magnetic field}}

In this section, we examine the magnetic-field dependence of the Raman scattering from the in-plane magnons of the AF phase ($M_1$ and $M_2$).

\begin{figure}[hbt!]
\centering
\includegraphics[width=0.84\textwidth]{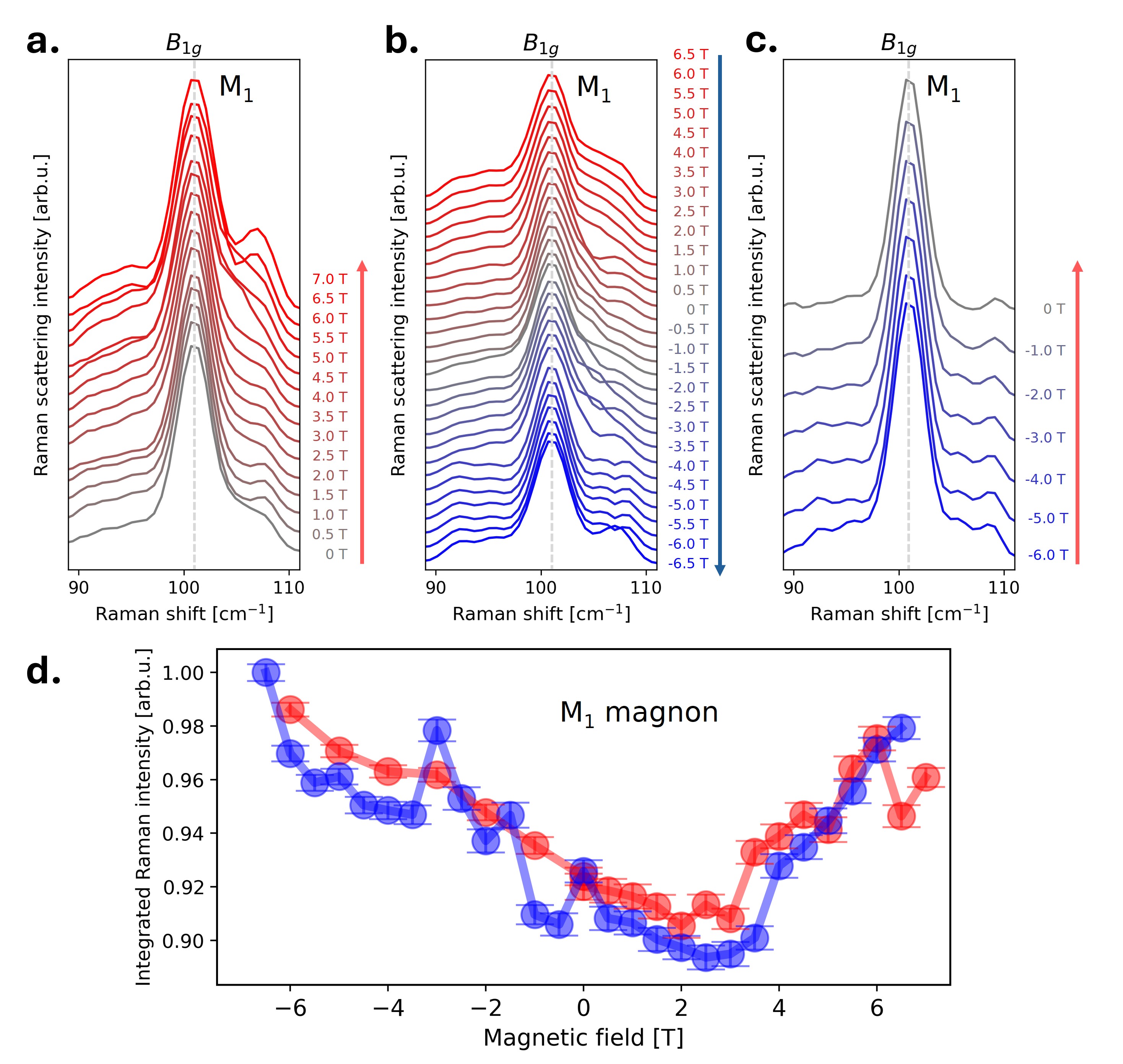}
\caption{\textbf{Evolution of the Raman scattering from the $M_1$ magnon in magnetic field.} Raman scattering from the $B_{1g}$ $M_1$ magnon of Ca$_2$RuO$_4$ measured as a function of magnetic field during \textbf{(a)} an increasing sweep, \textbf{(b)} a decreasing sweep, and \textbf{(c)} a closing sweep. The position of the $M_1$ magnon mode is indicated by a gray dashed line. \textbf{(d)} Magnetic-field dependence of the integrated Raman intensity of the $M_1$ mode (calculated as the sum of $B_{1g}$ and $A_g$ channels to avoid Faraday rotation artifacts), evaluated around 102 cm$^{-1}$ within a 5 cm$^{-1}$ window, for the field loop shown in Panels (a)–(c). In contrast to the $M^*$ mode, the integrated spectral weight of the $M_1$ magnon does not exhibit hysteresis. Red and blue arrows indicate opposite magnetic-field sweep directions. The integrated $M_1$ intensities are normalized to their maximum value across the entire magnetic-field range.}
\label{S2}
\end{figure}

Figure \ref{S2} demonstrates that the $M_1$ single-magnon mode is almost unaffected by the external magnetic field applied along the $c$-axis. Figure \ref{S2}(a-c) displays the $B_{1g}$ Raman spectra centered around the $M_1$ mode during the increasing (0 T to 7 T), decreasing (6.5 T to -6.5 T), and closing (-6.5 T to 0 T) magnetic-field sweeps. The magnon position and its Raman cross section remain almost unchanged. Figure \ref{S2}(d) quantifies the integrated Raman intensity of the $M_1$ mode across the full magnetic cycle. Unlike the $M^*$ mode, the $M_1$ intensity does not significantly change with a $c$-axis magnetic field and, crucially, does not exhibit any hysteresis.

\begin{figure}[hbt!]
\centering
\includegraphics[width=0.84\textwidth]{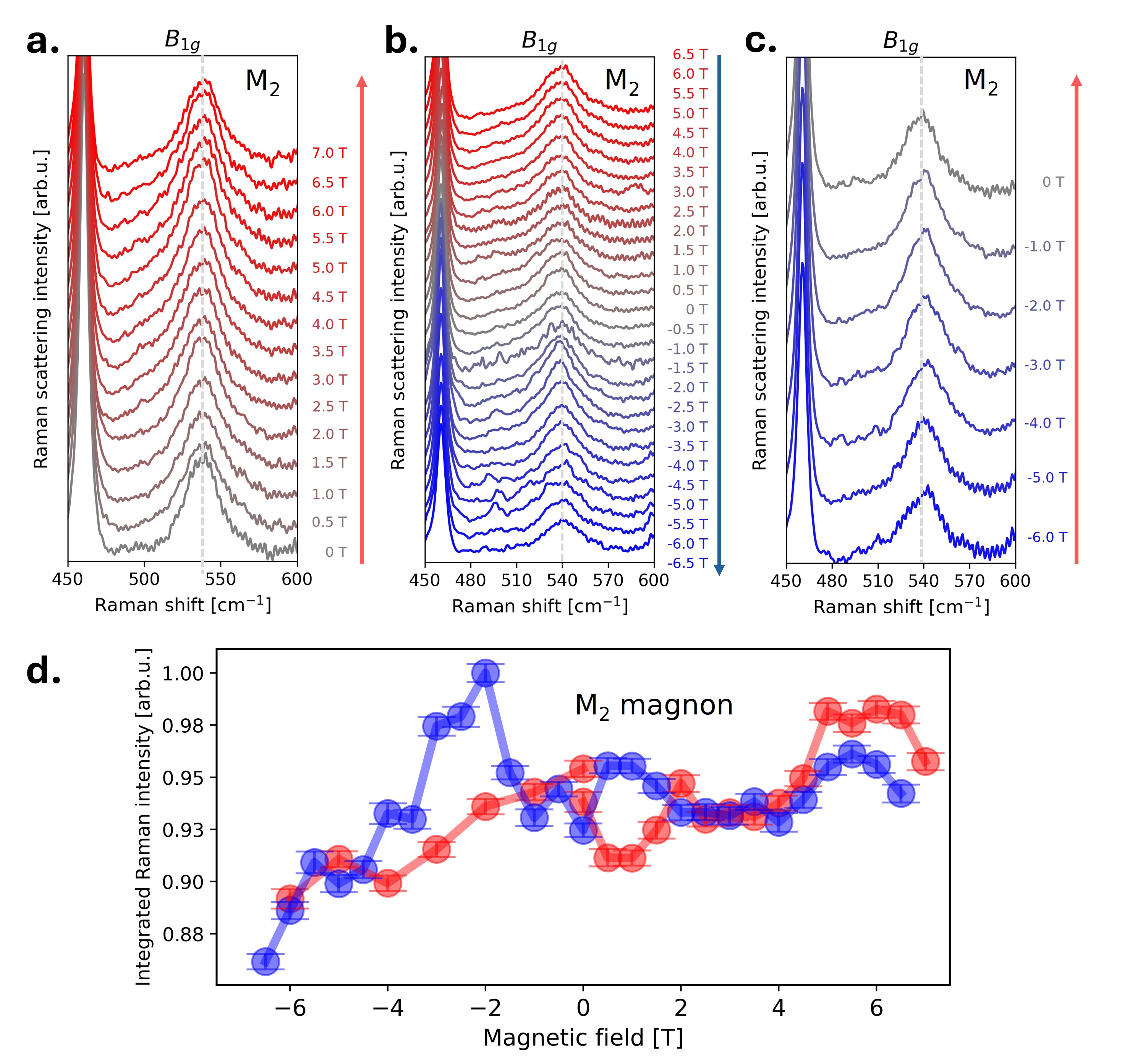}
\caption{\textbf{Evolution of the Raman scattering from the $M_2$ two-magnon in magnetic field.}
Raman scattering from the $B_{1g}$ $M_2$ two-magnon of Ca$_2$RuO$_4$ measured as a function of magnetic field during \textbf{(a)} an increasing sweep, \textbf{(b)} a decreasing sweep, and \textbf{(c)} a closing sweep. The position of the $M_2$ mode is indicated by a gray dashed line. \textbf{(d)} Magnetic-field dependence of the integrated Raman intensity of the $M_2$ mode (calculated as the sum of $B_{1g}$ and $A_g$ channels to avoid Faraday rotation artifacts), evaluated around 538 cm$^{-1}$ within a 30 cm$^{-1}$ window, for the field loop shown in Panels (a)-(c). In contrast to the $M^*$ mode, the integrated spectral weight of the $M_2$ two-magnon does not exhibit hysteresis. Red and blue arrows indicate opposite magnetic-field sweep directions. The integrated $M_2$ intensities are normalized to their maximum value across the entire magnetic-field range.}
\label{S3}
\end{figure}

Figure \ref{S3} shows the corresponding evolution of the higher-energy $M_2$ two-magnon broad excitation across the same field sweeps. Figure \ref{S3}(a-c) details the $B_{1g}$ Raman spectra in the  $M_2$ region as the field is varied across the loop. Figure \ref{S3}(d) confirms that the integrated spectral weight of the $M_2$ mode is almost constant and does not exhibit any hysteretic behavior.

Taken together, Figures \ref{S2} and \ref{S3} confirm that the primary in-plane antiferromagnetic order is not altered by the external $c$-axis magnetic field.

\newpage
\section{\protect\NoCaseChange{Integrated spectral weight of the $B_{1g}$ Raman phonons as a function of the magnetic field}}

\begin{figure}[hbt!]
\centering
\includegraphics[width=\textwidth]{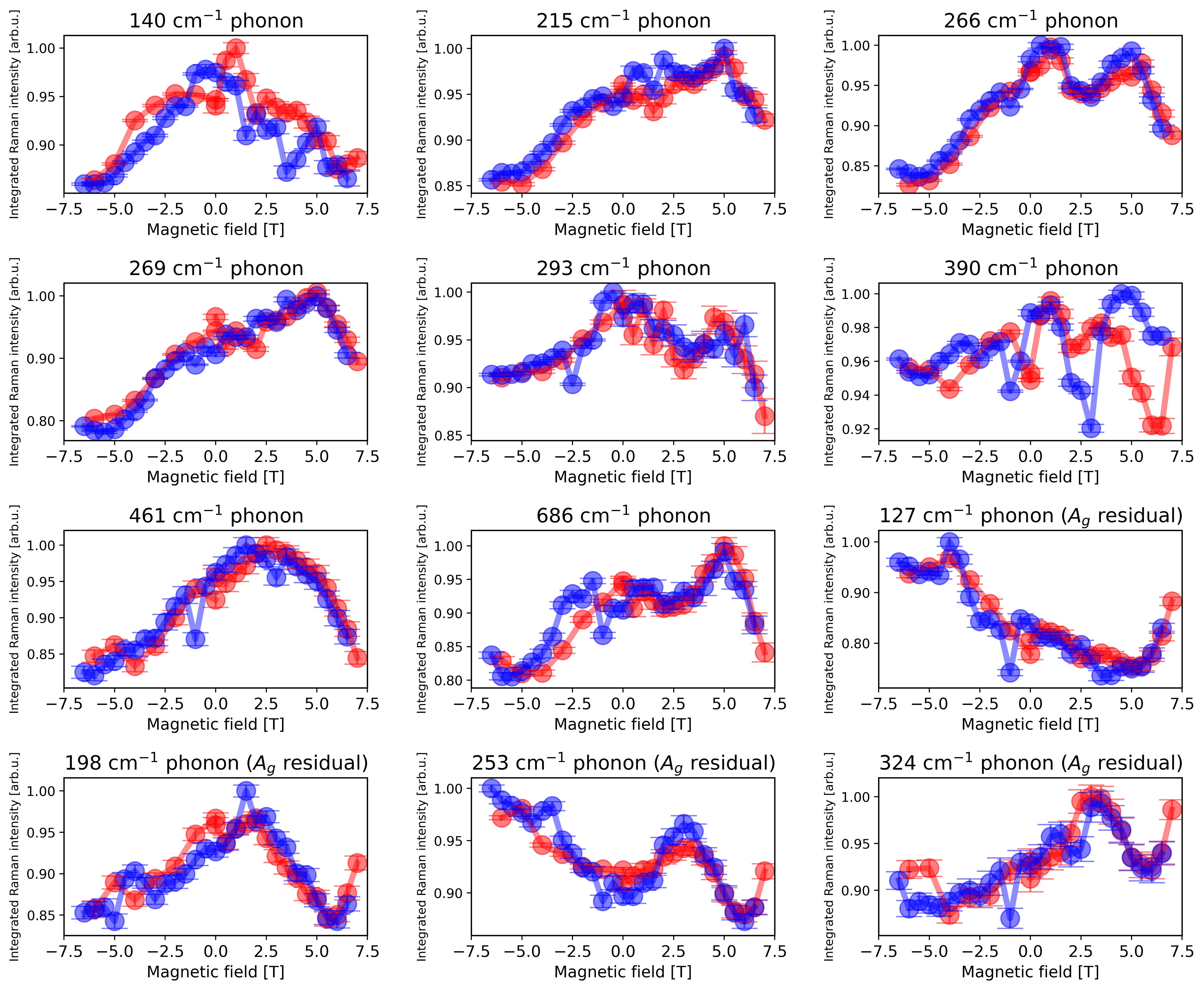}
\caption{\textbf{Integrated spectral weight of the $B_{1g}$ Raman phonons as a function of the magnetic field.}
Each panel shows the magnetic-field dependence of the integrated Raman intensity (calculated as the sum of $B_{1g}$ and $A_g$ channels to avoid Faraday rotation artifacts) for the $B_{1g}$ phonon modes of the AF phase. The residual $A_g$ modes at zero field are also included. The red and blue traces correspond to opposite magnetic-field sweep directions. No hysteresis is observed in the Raman spectral weight of any of the tracked phonon modes. The integrated intensities of each phonon are normalized to their maximum value across the entire magnetic-field range.}
\label{fig:S8}
\end{figure}

Figure \ref{fig:S8} shows the integrated spectral weights of the $B_{1g}$ phonons in the antiferromagnetic phase (1.6~K) as a function of the magnetic field. Each panel plots the integrated Raman scattering of a single lattice mode during the full forward and backward magnetic-field sweeps. To further avoid artifacts from Faraday rotation at high magnetic fields, the plotted values are calculated as the sum of the $B_{1g}$ and $A_g$ components ($I_{\text{Tot}} = I_{B_{1g}} + I_{A_g}$). These panels also display the residual $A_g$ symmetry modes (at 127, 198, 253, and 324~cm$^{-1}$, see Figure \ref{S1} of the SM) that appear in the $B_{1g}$ channel at zero field. For every tracked mode, the red (increasing) and blue (decreasing) traces almost overlap and do not significantly change across the field range. The results demonstrate an absence of hysteresis for each vibrational mode and confirm that all the $B_{1g}$ phonons are not affected by the applied out-of-plane magnetic field.

\clearpage
\newpage

\section{\protect\NoCaseChange{Spectral redistribution of the $H$ mode in magnetic field}}

\begin{figure}[hbt!]
\centering
\includegraphics[width=0.95\textwidth]{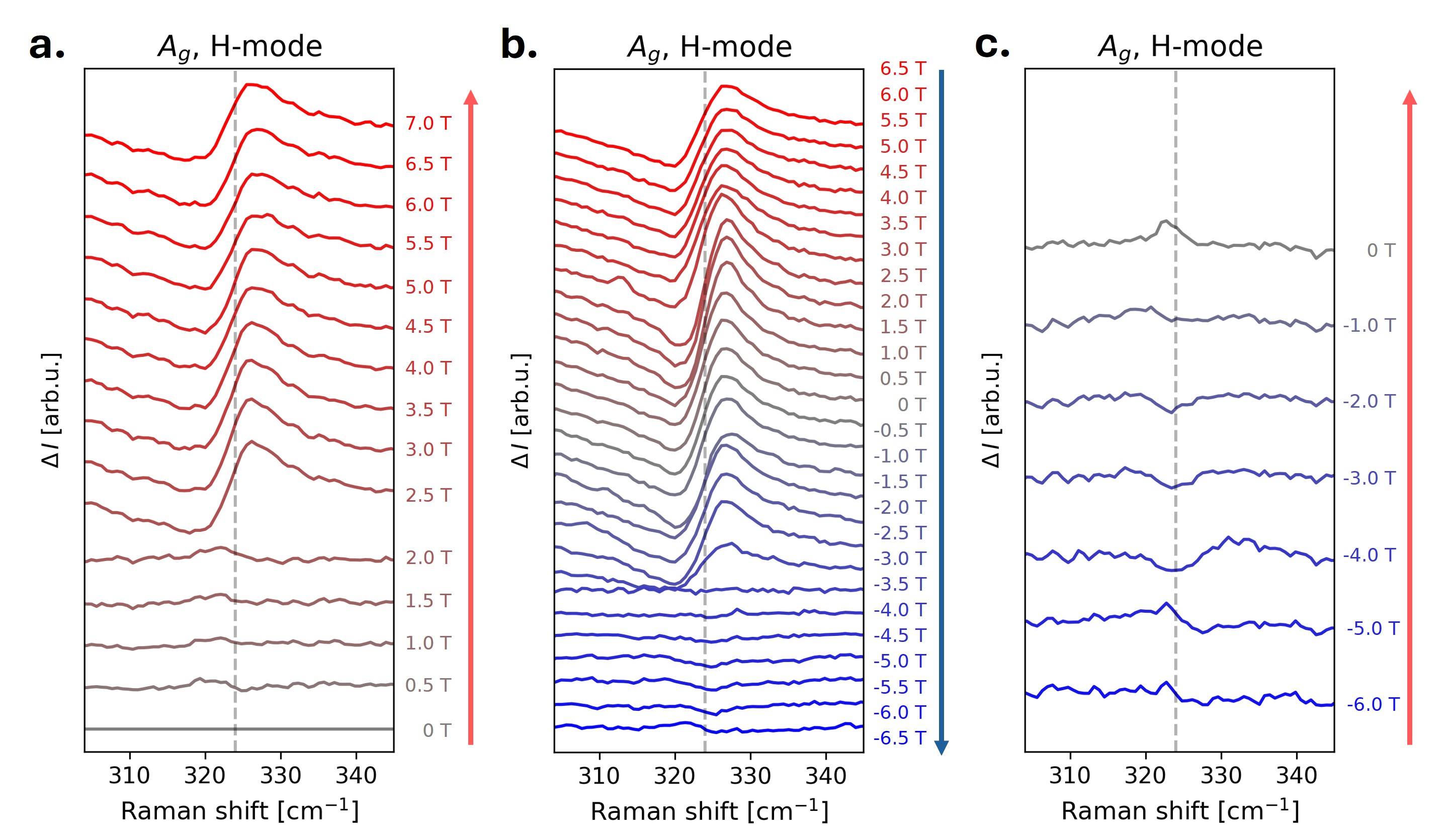}
\caption{\textbf{Spectral redistribution of the $H$ mode in magnetic field with respect to the zero-field spectrum.}
Raman $A_{g}$ spectra at 1.6 K in magnetic field applied along the $c$-axis, plotted as $\Delta I(B)=I(B)-I(0~\mathrm{T}_{\uparrow})$, where $I(0~\mathrm{T}_{\uparrow})$ is the spectrum measured at 0 T at the beginning of the magnetic-field loop, corresponding to the 0 T spectrum of Figure \ref{fig:fig3}(a) (left panel). The magnetic-field hysteresis loop is shown as a sequence of sweeps: \textbf{(a)} increasing field from 0 T to 7 T, \textbf{(b)} decreasing field from +6.5 T to $-6.5$ T, and \textbf{(c)} a closing sweep from $-6.5$ T back to 0 T. The vertical dashed line marks the frequency of the $H$ mode (324 cm$^{-1}$).}
\label{fig:S4}
\end{figure}

Figure \ref{fig:S4} shows the magnetic field-induced variations in the $A_g$ Raman spectrum around the $H$ phonon frequency, plotted as the differential intensity $\Delta I(B) = I(B) - I(0~\mathrm{T}_{\uparrow})$ relative to the zero-field reference. By subtracting the zero-field response, this differential signal reveals a hysteretic reconstruction of the underlying Higgs mode. Through magneto-elastic coupling, this spectral redistribution of the Higgs mode is directly mapped onto the $H$ phonon. Consequently, it drives the hysteretic modification of the Fano parameter of the $H$ mode, as quantified in the main text (Figure \ref{fig:fig3}).

\clearpage
\newpage

\section{\protect\NoCaseChange{Fano-lineshape fits of the $H$ mode in magnetic field}}

\begin{figure}[hbt!]
\centering
\includegraphics[width=\textwidth]{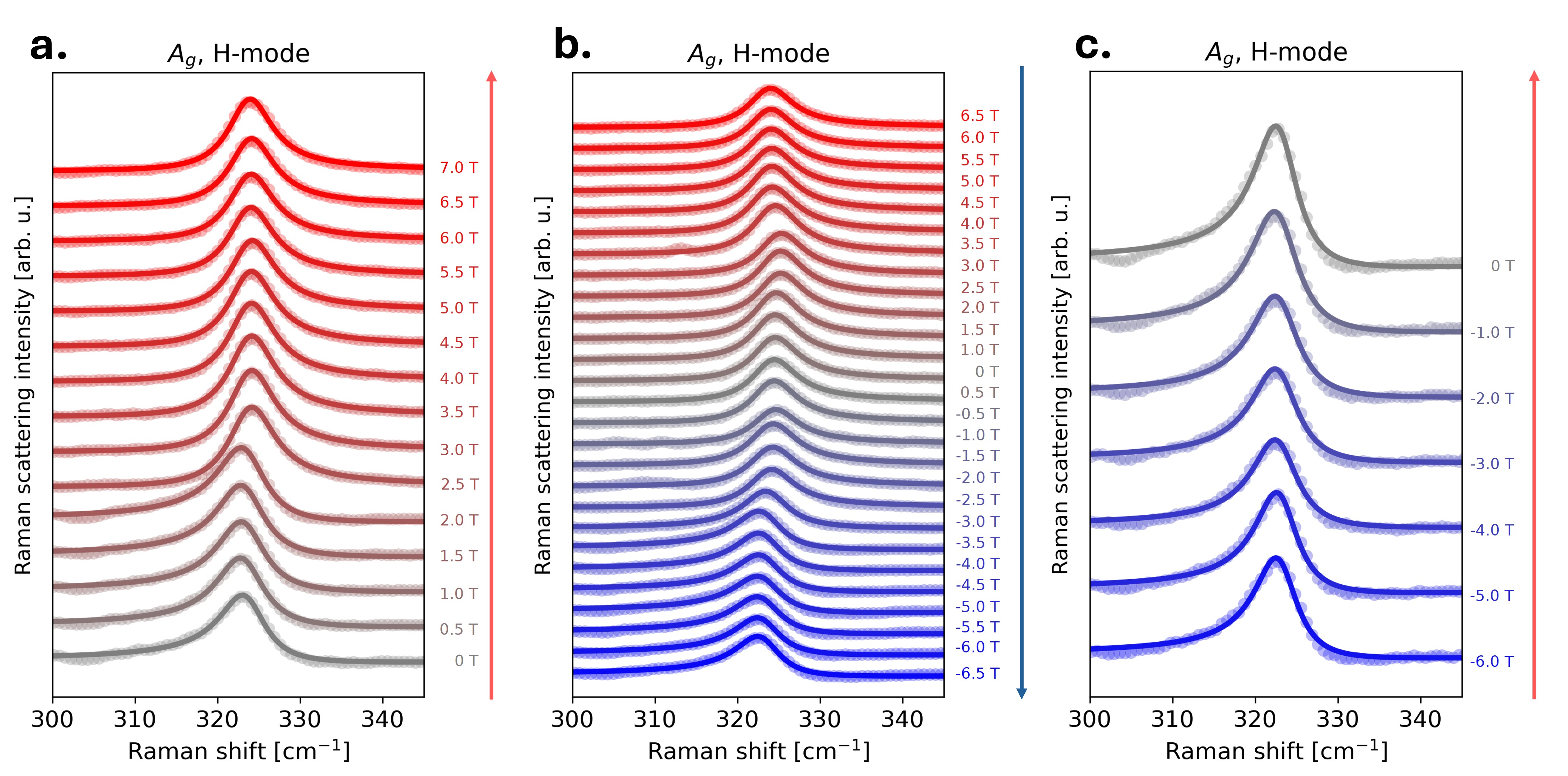}
\caption{\textbf{Fano-lineshape fits of the $A_{g}$-symmetry $H$ mode as a function of the magnetic field.}
Panels \textbf{(a)}, \textbf{(b)}, and \textbf{(c)} correspond to the increasing, decreasing, and closing field sweep directions, respectively. Solid lines denote fits of the $H$ mode with a Fano profile, while symbols represent the raw Raman spectra.}
\label{fig:S5}
\end{figure}

Figure \ref{fig:S5}(a-c) reports the raw $A_g$ Raman spectra of the $H$ phonon alongside their Fano fits across the magnetic-field cycle. The fits capture the transition from negative to positive Fano asymmetry, confirming the hysteretic behavior presented in Figure \ref{fig:fig3}.

\clearpage
\newpage

\section{\protect\NoCaseChange{Fano-lineshape fits of an $A_g$ Raman mode off-resonance with the Higgs mode}}

\begin{figure}[hbt!]
\centering
\includegraphics[width=\textwidth]{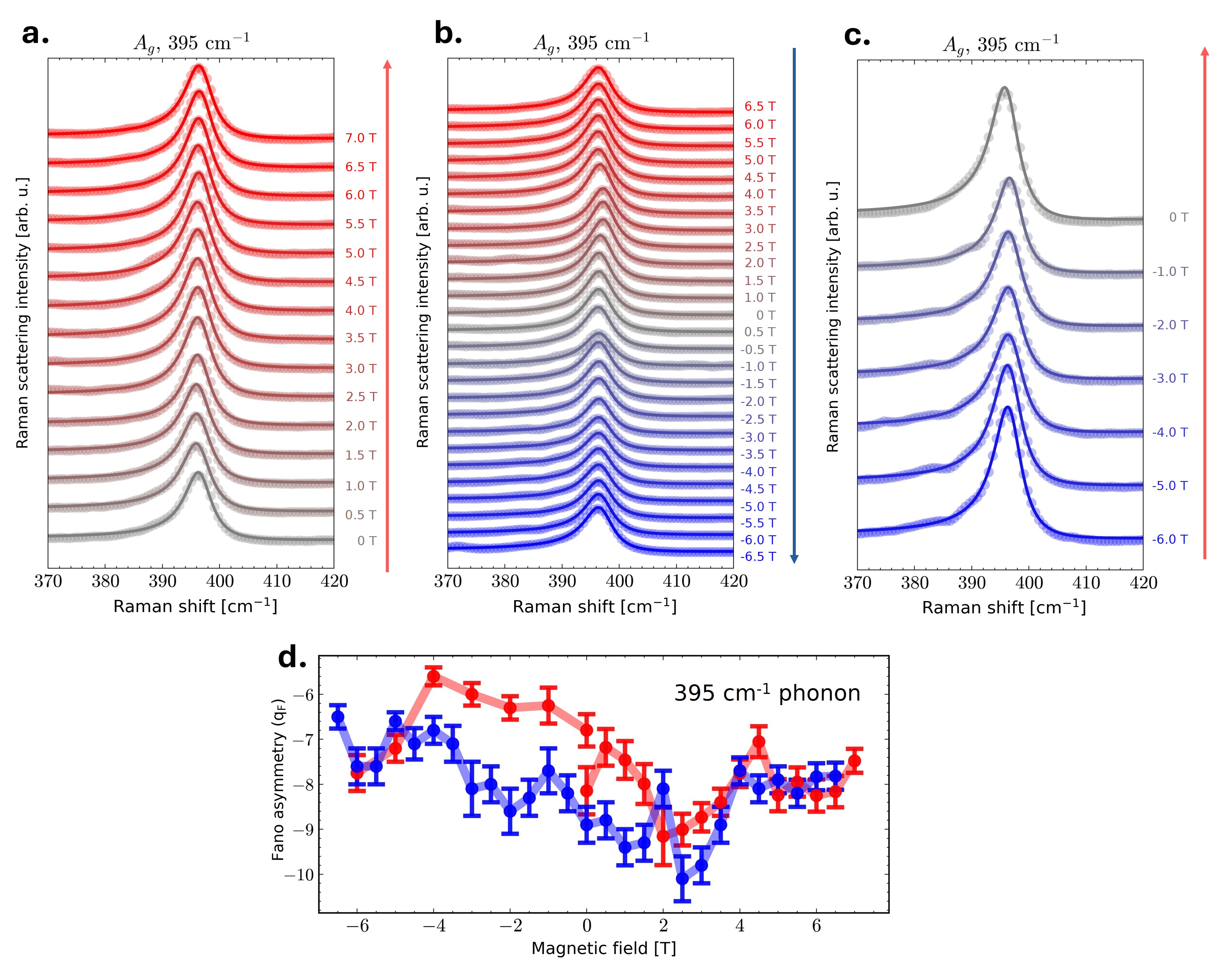}
\caption{\textbf{Fano-lineshape fits of the $A_{g}$-symmetry phonon at 395 cm$^{-1}$ and evolution of its asymmetry parameter in magnetic field.}
Panels \textbf{(a)}, \textbf{(b)}, and \textbf{(c)} show the $A_g$ Raman spectra (symbols) and corresponding Fano profile fits (solid lines) for the increasing, decreasing, and closing field sweep directions, respectively. The spectra are obtained after subtracting the residual contribution of a $B_{1g}$ symmetry mode centered around 390 cm$^{-1}$ (fitted by a Gaussian profile), which emerges at high magnetic fields due to residual Faraday rotation. \textbf{(d)} Magnetic-field dependence of the Fano asymmetry parameter $q_{\mathrm{F}}$ of the 395 cm$^{-1}$ phonon, which exhibits no sign change or hysteretic behaviour, in contrast to the $H$ mode.}
\label{fig:S_Ag_phon_fits}
\end{figure}

To verify that the hysteretic behavior of the $H$ mode is driven by its magneto-elastic coupling with the resonant Higgs excitation, we track another $A_g$ symmetry phonon at 395 cm$^{-1}$, which is out-of-resonance with the Higgs peak identified in~\cite{souliou2017_supp}. Figure \ref{fig:S_Ag_phon_fits}(a-c) presents the $A_g$ Raman spectra of this selected phonon mode plotted alongside their respective Fano lineshape fits. The spectra are obtained by subtracting a residual $B_{1g}$ symmetry mode around 390 cm$^{-1}$, which emerges at high magnetic fields due to residual Faraday rotation. This spurious contribution is fitted with a Gaussian profile and removed from the $A_g$ Raman spectra. From these fits, we extract the Fano asymmetry parameter $q_{\mathrm{F}}$ for the 395 cm$^{-1}$ $A_g$ phonon.

As shown in Figure \ref{fig:S_Ag_phon_fits}(d), its field dependence exhibits no sign change or hysteretic behavior. This suggests that the Fano anomaly of the $H$ mode is not a general feature shared by other $A_g$ phonons, but is linked to its magneto-elastic coupling with the underlying Higgs mode.

\clearpage
\newpage

\section{\protect\NoCaseChange{Correction of the Raman spectra for Faraday rotation in the microscope objective}}

\begin{figure}[hbt!]
\centering
\includegraphics[width=0.95\textwidth]{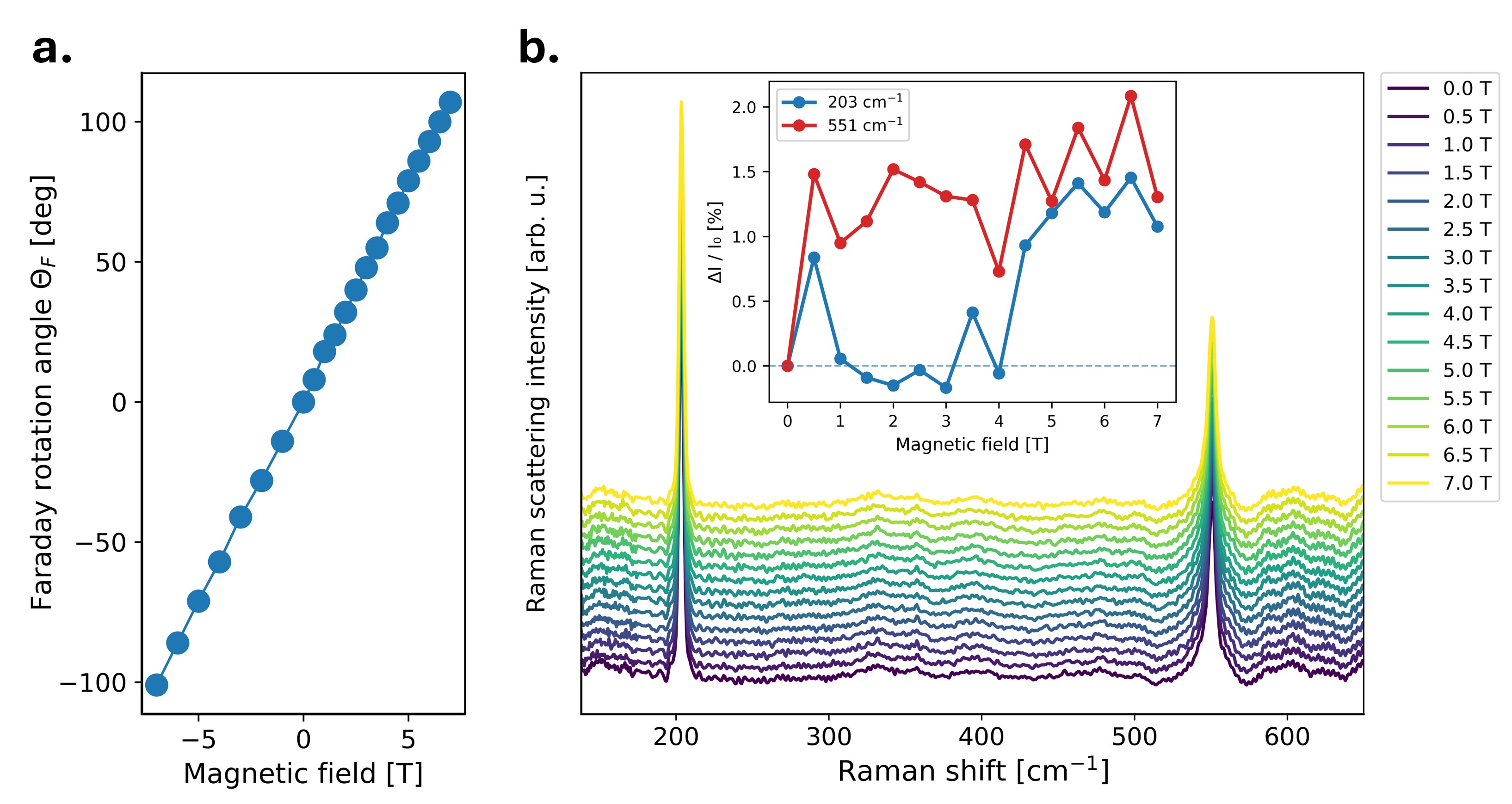}
\caption{\textbf{Correction of the Raman spectra for Faraday rotation in the objective.}
\textbf{(a)} Dependence of the polarization rotation of the scattered light induced by Faraday effect in the optical elements of the microscope objective ($\theta_F(B)$). For this calibration, elastically scattered light from the non-magnetic material Sr$_2$RuO$_4$ is collected. \textbf{(b)} Magnetic-field dependence of the Raman spectra of the non-magnetic compound Sr$_2$RuO$_4$ at 300 K after applying the Faraday-rotation correction protocol. The inset shows the integrated intensities of the 203 and 551 cm$^{-1}$ $A_g$ phonons, confirming the effectiveness of the correction.}
\label{fig:S7}
\end{figure}

Figure \ref{fig:S7} details the protocol used to correct the spurious polarization rotation caused by the Faraday effect in the microscope objective in a magnetic field. As light travels through the objective lenses parallel to the external field, its polarization rotates by an angle $\theta_F(B)$. This field-induced rotation unavoidably mixes the $A_g$ and $B_{1g}$ channels in the raw spectra, leading to apparent field-dependent variations of the Raman modes. 

To systematically disentangle the intrinsic Raman signals, Figure \ref{fig:S7}(a) reports the polarization rotation angle $\theta_F(B)$ measured on a non-magnetic $\text{Sr}_2\text{RuO}_4$ reference sample as a function of the magnetic field. Using this calibration curve, we re-rotate both the incident and scattered polarization fields by $-\theta_F(B)/2$. This correction compensates for the Faraday effect, ensuring that the incident polarization on the sample remains constant in magnetic fields, and that the analyzed light corresponds to the pure parallel ($A_g$) or crossed ($B_{1g}$) polarization geometries.

Figure \ref{fig:S7}(b) plots the integrated intensities of two vibrational Raman modes in $\text{Sr}_2\text{RuO}_4$ (at 203~$\text{cm}^{-1}$ and 551~$\text{cm}^{-1}$). For such a non-magnetic sample, no $B$ field dependence of the mode intensities is expected. These modes are $A_g$ phonons associated with the $\text{RuO}_6$ octahedra; therefore, for a fixed incident polarization, they must remain maximized under parallel detection and vanish in extinction detection. The observation that their corrected Raman intensities display a field-independent behavior confirms the effectiveness of the protocol used to correct the Faraday rotation inside the objective.

However, this correction alone does not account for the difference in the Verdet coefficient between $\text{Ca}_2\text{RuO}_4$ and $\text{Sr}_2\text{RuO}_4$. To rule out any artifacts arising from this difference, especially at high magnetic fields, when tracking the field dependence of the mode intensities in all the figures we always plot the total sum of the two polarization channels ($I_{\text{Tot}} = I_{A_g} + I_{B_{1g}}$). Plotting this sum clearly shows the emergence of new modes, given also the corepresentation analysis of Section \ref{section:S2} showing a mixing of the channels in a magnetic field. 

\clearpage
\newpage

\section{\protect\NoCaseChange{Comparison of $B_{1g}$ and $A_g$ Raman spectra at zero and finite magnetic fields}}

\begin{figure}[hbt!]
\centering
\includegraphics[width=0.75\textwidth]{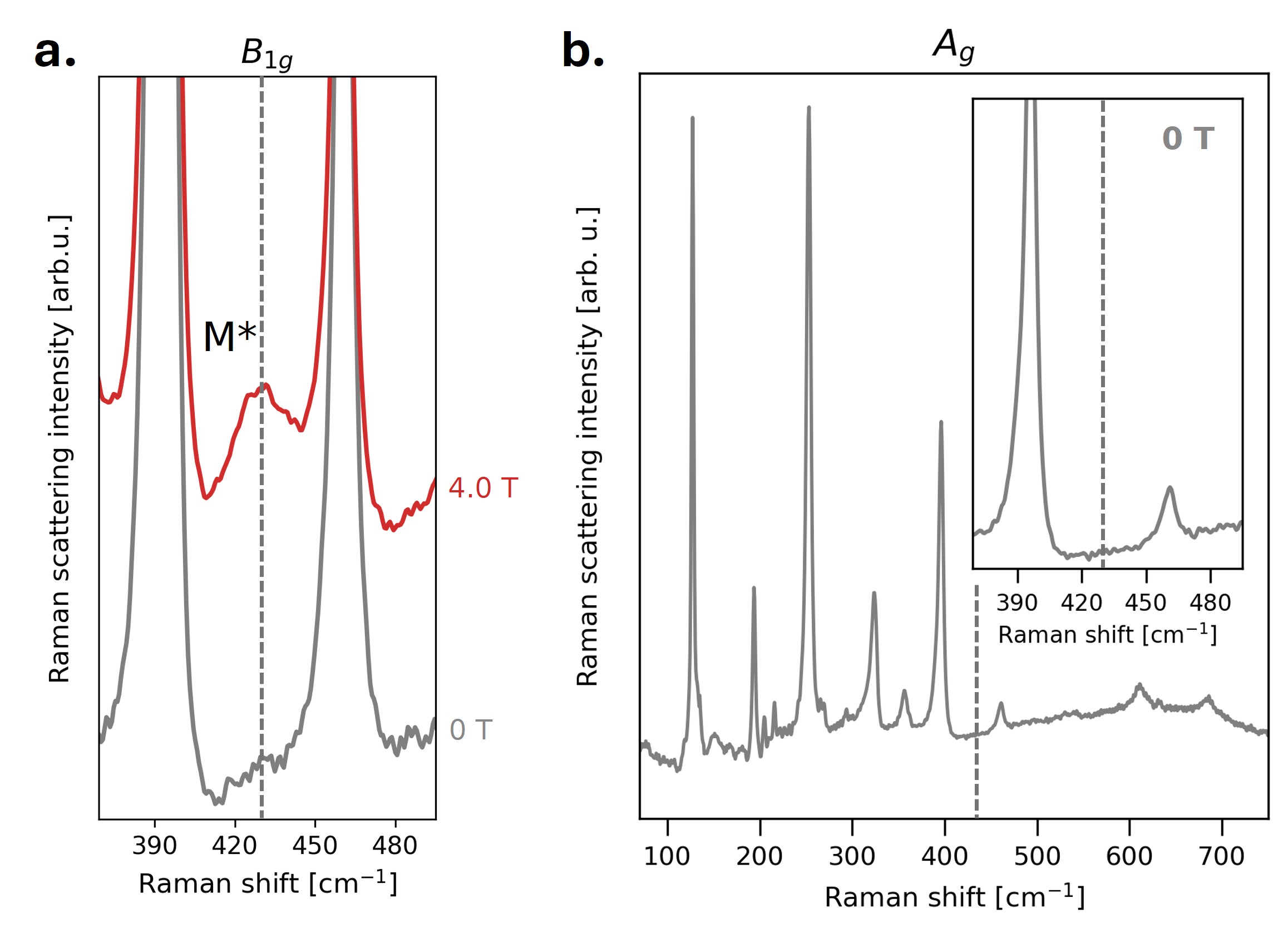}
\caption{\textbf{Comparison of $B_{1g}$ Raman spectra in magnetic field and $A_g$ Raman spectrum at zero-field.}
\textbf{(a)} $B_{1g}$ Raman spectra measured at $B=0$ T and $B=4$ T at 1.6 K, showing the emergence of the $M^*$ mode for $B>B_{c1}$. \textbf{(b)} $A_g$ Raman spectrum at zero magnetic field; the inset shows a zoom around the frequency of the $M^*$ mode. The frequency of the $M^*$ mode observed in $B_{1g}$ (432 cm$^{-1}$) is indicated by the gray dashed line. No Raman mode is present in $A_g$ at 0 T at the $M^*$ frequency.}
\label{fig:S6}
\end{figure}

To prove that the $M^*$ mode is a new excitation activated by the magnetic field and does not originate from polarization leakage, we compare in Figure \ref{fig:S6} the $B_{1g}$ spectrum above the critical field with the $A_g$ spectrum at zero field. Figure \ref{fig:S6}(a) shows the $B_{1g}$ spectrum at 4~T (above $B_{c1}$), demonstrating the activation of the $M^*$ mode at 432~cm$^{-1}$. Figure \ref{fig:S6}(b) reports the $A_g$ spectrum at zero field ($B = 0$~T). The inset provides a magnified view of the frequency region around 432~cm$^{-1}$, showing a complete absence of scattering intensity in the $A_g$ channel. Because there is no $A_g$ signal at this specific frequency, the $M^*$ peak cannot arise from a spurious mixing of the two channels caused by Faraday rotation. This suggests that the $M^*$ excitation is a new mode, forbidden by the selection rules at zero magnetic field, and that it exhibits $B_{1g}$ symmetry once activated above $B_{c1}$.

\clearpage
\newpage

\section{\protect\NoCaseChange{Effective free energy for dipolar and quadrupolar sectors}}
\label{sec:free_energy}
In this section we present a phenomenological derivation of the free energy $\mathcal{F}_q$ for the quadrupolar order parameter $q\equiv Q^{xz}=<S^x S^z+S^z S^x>$, and derive the additional contribution arising from the inclusion of an external magnetic field.

The quadrupolar operator is a bilinear combination of spin operators and is therefore naturally invariant under time-reversal symmetry. As a consequence, the most general Landau expansion for the quadrupolar order parameter admits both even and odd powers of $q$. Up to fourth order in $q$, the most general free energy reads
\begin{equation}
\mathcal{F}_q=\frac{r}{2}q^2-\frac{s}{3}q^3
+\frac{u}{4}q^4.
\label{Fqsup}
\end{equation}
The absence of a linear term follows from the fact that it can always be removed by a constant shift of $q$, while stability requires $u>0$. The stationary points are determined by $\partial_q\mathcal{F}_q=0$. In the regime of interest, with $r>0$, $s>0$, and $s^2>4ru$, $\mathcal{F}_q$ exhibits two minima, at $q=0$ and $q=q^*$, with $q^*\equiv\left( s+\sqrt{s^2-4ru} \right)/(2u)$, separated by a local maximum at $q=\left( s-\sqrt{s^2-4ru} \right)/(2u)$. 

We now discuss the coupling between dipolar and quadrupolar sectors, the role of the external magnetic field, and the resulting effective free energy for the quadrupolar order parameter. The bare free energy for the dipolar order parameter, defined as $n\equiv\sqrt{n_x^2+n_z^2}$, with $n_x$ and $n_z$ the in-plane and canted components of the staggered magnetization, respectively, is given by
\begin{equation}
\mathcal{F}_n=-\frac{r_n}{2}n^2+
\frac{u_n}{4}n^4,
\label{Fn}
\end{equation}
with $r_n,u_n>0$. $\partial_n\mathcal{F}_n=0$ yields a minimum at finite staggered moment $\overline{n}\equiv\sqrt{r_n/u_n}$. Regarding the coupling between dipolar and quadrupolar sectors and the applied field, symmetry allows for the trilinear terms $\lambda Q^{\alpha\beta}n^\alpha n^\beta$ and $\mu Q^{\alpha\beta}n^\alpha B^\beta$. Since the quadrupolar tensor is traceless ($Q^{\alpha\alpha}=0$) and $\mathbf{B}=B\mathbf{z}$, the coupling sector reduces to $\mathcal{F}_{nq}=\lambda q n_x n_z+\mu q n_x B$. Introducing the canting angle $\theta$, such that $n_x=n\cos\theta$ and $n_z=n\sin\theta$, $\mathcal{F}_{nq}$ reduces, at leading order in $\theta$, to Eq.\ \ref{Fnq} of the main text. To derive the effective free energy in terms of $q$ only, we first solve $\partial_n\left(\mathcal
F_n+\mathcal{F}_{nq}\right)=0$, i.e.,
\begin{equation}
u_n n^3-r_n n+
2\lambda\theta n q+
\mu B q=0
\end{equation}
for $n$. Up to second order in $q$, and to leading order in the external field and coupling constants, we obtain
\begin{equation}
n=\overline{n}-
\left(
\frac{\lambda\theta}{\sqrt{r_n u_n}}+\frac{B\mu}{2 r_n}
\right)q
-\frac{B\mu\lambda\theta }
{r_n^2}q^2.
\label{nper}
\end{equation}
Substituting Eq.\ \ref{nper} into the total free energy $\mathcal{F}_q+\mathcal{F}_n+\mathcal{F}_{nq}$, we get the effective free energy
\begin{equation}
\tilde{\mathcal{F}}_q=\frac{\tilde{r}}{2}q^2-\frac{s}{3}q^3+\frac{u}{4}q^4,
\end{equation}
where
\begin{equation}
\tilde{r}=r-\kappa B,
\end{equation}
and $\kappa\equiv 2\lambda \mu\theta/\sqrt{r_n u_n}$, as defined in the main text. For $-B_{c2}<B<B_{c1}$, with $B_{c1}\equiv r/\kappa$ and $B_{c2}\equiv s^2/(4u\kappa)-r/\kappa$, the renormalized free energy $\tilde{\mathcal{F}}_q$ exhibits two minima at $q=0$ and $q=\tilde{q}^*$, where $\tilde{q}^*\equiv\left( s+\sqrt{s^2-4\tilde{r}u} \right)/(2u)$. For $B\geq B_{c1}$, the condition $\partial_q^2\mathcal{F}_{q=0}\leq 0$ implies that $q=0$ becomes unstable, leaving $q=\tilde{q}^*$ as the only minimum. Conversely, for $B\leq-B_{c2}$ the condition
$\partial_q^2\mathcal{F}_{q=\tilde{q}^*}\leq0$ renders $\tilde{q}^*$ unstable, leaving $q=0$ as the only minimum.

\clearpage
\newpage

\section{\protect\NoCaseChange{Atomic displacement of the $H$ phonon mode}}

\begin{figure}[htbp]
    \centering
    \includegraphics[width=0.6\columnwidth]{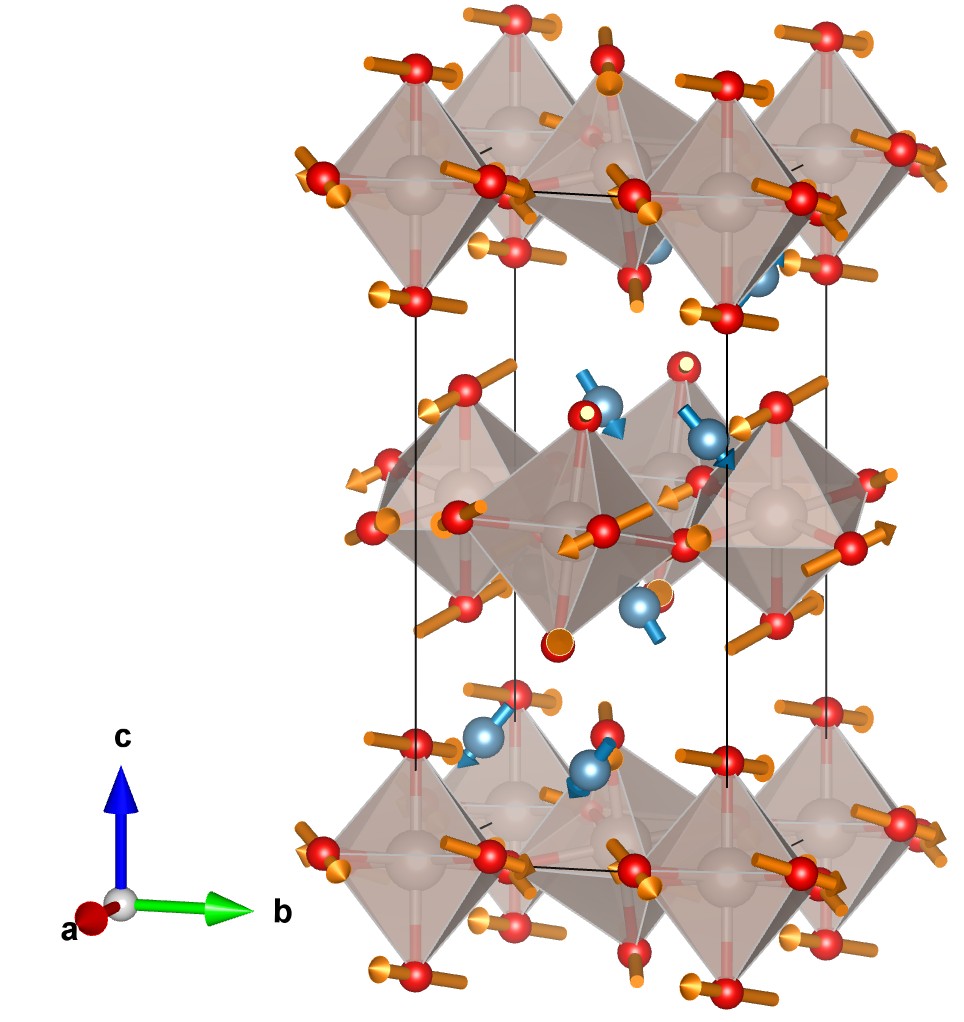} 
    \caption{Sketch of the atomic displacements for the $H$ phonon mode at $324\,\mathrm{cm}^{-1}$, derived from DFT calculations. The orange and light-blue arrows indicate the atomic displacements.}
    \label{fig:supp_Hmode}
\end{figure}

To visualize the lattice dynamics associated with the $H$ phonon mode at $324\,\mathrm{cm}^{-1}$, we extracted the corresponding atomic displacements using the density-functional theory (DFT) calculations detailed in Section~\ref{section:S2}. The $H$ mode displacements are presented in Figure~\ref{fig:supp_Hmode}.

\clearpage

\vspace{1cm}
\begin{center}
    \textbf{\large Supplementary References}
\end{center}

\end{document}